\documentclass[letterpaper, useASM,usenatbib]{mn2e}

\usepackage{rotating}
\usepackage{amssymb,amsmath}	
\usepackage{natbib}
\usepackage{graphicx}
\usepackage{grffile}
\usepackage{mathrsfs}
\usepackage{color}
\usepackage[normalem]{ulem}		
\usepackage{cancel}			
\usepackage{times}
\usepackage[paperwidth=9.0in, paperheight=11.5in]{geometry}
\usepackage{fullpage}
\usepackage{verbatim}

\newcommand{\be}{\begin{eqnarray}}
\newcommand{\ee}{\end{eqnarray}}
\newcommand{\ba}{\be}
\newcommand{\ea}{\ee}

\newcommand{\Msun}{M_{\odot}}
\newcommand{\Edot}{\dot E}
\newcommand{\dcrab}{d_{\rm Crab}}
\newcommand{\dmax}{d_{\rm max}}
\def\Dgpc{d_{\rm Gpc}}

\newcommand{\Snu}{S_{\nu}}
\newcommand{\Snumax}{S_{\rm \nu, max}}
\newcommand{\Snumin}{S_{\rm \nu, min}}

\newcommand{\Evec}{{\bf E}}
\newcommand{\Bvec}{{\bf B}}
\newcommand{\Omvec}{\mathbf \Omega}
\newcommand{\rlc}{r_{\rm lc}}
\newcommand{\ngj}{n_{\rm GJ}}

\def\msun{M_\odot}
\def\rsun{R_\odot}

\newcommand{\Br}{\Delta\nu_r}		
\newcommand{\tp}{t^{\prime}}

\newcommand{\nhat}{{\mathbf{\hat n}}}

\newcommand{\sinc}{ {\rm sinc}}

\def\kms{\,{\rm km\,s^{-1}}}
\def\Dnud{\Delta\nu_r}
\def\gint{\gamma_{\rm int}}

\newcommand\epse{\epsilon_t}

\def\dte{\delta t_e}
\def\nuunit{\nu_{\rm GHz}}

\def\te{t_e}

\def\tec{t_{e,c}}
\def\dtheta{\Delta\theta}
\def\dphi{\Delta\phi}
\def\dr{\Delta r}
\def\drl{\Delta R_l}
\def\eps{\epsilon}
\def\that{{\mbox{\boldmath ${\hat t}$}}}
\def\chat{{\mbox{\boldmath ${\hat c}$}}}
\def\rhat{{\mbox{\boldmath ${\hat r}$}}}

\def\phihat{{\mbox{\boldmath ${\hat\phi}$}}}
\def\rvec{{\mbox{\boldmath $r$}}}
\def\zhat{{\mbox{\boldmath $\hat z$}}}
\def\xhat{{\mbox{\boldmath $\hat x$}}}
\def\yhat{{\mbox{\boldmath $\hat y$}}}
\def\vvec{{\mbox{\boldmath $v$}}}
\def\avec{{\mbox{\boldmath $a$}}}
\def\dotprod{{\mbox{\boldmath $\cdot$}}}
\def\crossprod{{\mbox{\boldmath $\times$}}}

\def\omco{\omega_{\rm coh}}
\def\nuco{\nu_{\rm coh}}
\def\nucounit{\nu_{\rm coh,GHz}}
\def\rlc{r_{\rm lc}}
\def\lhat{{\mbox{\boldmath $\hat\ell$}}}
\def\rtil{{\tilde r}}
\def\temc{\te-\tec}
\def\omtemc{\Omega(\temc)}
\def\ompk{\omega_{\rm pk}}
\def\nupk{\nu_{\rm pk}}
\def\gamco{\gamma_{\rm coh}}
\def\gaminco{\gamma_{\rm incoh}}
\def\Fco{F(\omega/\omco)}
\def\Gco{C(\omega/\omco)}		
\def\Dtec{\Delta\tec}

\newcommand{\zp}{z^{\,\prime}}

\newcommand{\Lnumin}{L_{\nu, \rm min}}

\newcommand{\NNS}{N_{\rm ns}}

\newcommand{\Birthrate}{{\Gamma_{\rm ns}}}
\newcommand{\BRm}{\dot n_{\rm ns, M} }
\newcommand{\BRminline}{\dot n_{\rm ns, M} }
\newcommand{\BRminlinezero}{\dot n_{\rm ns, M,0} }

\newcommand{\BRmFiducial}{\dot n_{\rm ns, M, -13}}

\newcommand{\NGone}{N_{b}}
\newcommand{\zmax}{z_{\rm max}}

\def\ratepermass{\BRminline}

\def\rhostar{\rho_{\star}(z)}
\def\rhostarnow{\rho_\star(0)}

\def\nuns{\nu_{ns}(z)}
\def\nunszero{\nu_{ns}(0)}
\def\nuhns{{\hat\nu}_{ns}(z)}

\def\psihat{{\widehat\psi}(z)}
\def\psihatp{{\widehat\psi}(\zp)}

\def\rtil{{\tilde r}}

\newcommand{\OmstarFiducial}{\Omega_{*, 0.003}}

\newcommand{\burstrate}{\Gamma_b}

\newcommand{\burstrateobs}{\Gamma_{\rm b,obs}}

\newcommand{\Vc}{\mathscr{V}_c}
\newcommand{\Vi}{\mathscr{V}_i}
\newcommand{\snu}{{f(\nu)}}

\newcommand{\burstrateH}{\Gamma_{\rm b, H}}

\newcommand{\onehalf}{\frac{1}{2}}

 \newcommand{\etagj}{\eta_{\rm GJ}}
 \newcommand{\sshotz}{s_{\nuz}}
\newcommand{\sshotzmax}{s_{\nuz, {\rm max}}}
\newcommand{\sshotmax}{s_{\nu, {\rm max}}}

\newcommand{\ashotz}{A_{\nuz}}

\def\aap{AAp}
\def\apj{ApJ}
\def\apjl{ApJL}
\def\mnras{MNRAS}

\def\nat{Nature}
\def\araa{ARAA}
\def\pre{Phys. Rev. E}
\def\prd{Phys. Rev. D}

\onecolumn

\begin{document}

\title[Supergiant Pulses from Extragalactic Neutron Stars]{Supergiant Pulses from Extragalactic Neutron Stars}
\author[J. M. Cordes \& Ira Wasserman]
{
	J. M. Cordes \thanks{E-mail: jmc33@cornell.edu} 
	\& Ira Wasserman
	\thanks{E-mail: ira@astro.cornell.edu} \\
 Astronomy Department, Cornell University, Ithaca, NY 14853}
\date{\today}

\maketitle

\Large

\label{firstpage}

\begin{abstract}
 We consider radio bursts that originate from extragalactic neutron stars
(NSs) by addressing three questions about source distances. What are the
physical limitations on coherent radiation at GHz frequencies? Do they permit
detection at cosmological distances? How many bursts per NS are needed to
produce the inferred burst rate $\sim 10^3$-$10^4 $sky$^{-1}$ day$^{-1}$? The
burst rate is comparable to the NS formation rate in a Hubble volume, requiring
only one per NS if they are bright enough. However, radiation physics causes us
to favor a closer population. More bursts per NS are then required but repeats
in 10 to 100 yr could still be negligible. Bursts are modeled as sub-ns,
coherent shot pulses superposed incoherently to produce ms-duration $\sim 1$ Jy
amplitudes; each shot-pulse can be much weaker than the burst amplitude,
placing less restrictive requirements on the emission process. Nonetheless,
single shot pulses are similar to the extreme, unresolved ($< 0.4$ ns) MJy shot
pulse seen from the Crab pulsar, which is consistent with coherent curvature
radiation emitted near the light cylinder by an almost neutral clump with net
charge $\sim \pm 10^{21}e$ and total energy $\gtrsim 10^{23}$ ergs. Bursts from
Gpc distances require incoherent superposition of $\sim 10^{12}d_{\rm Gpc}^2$
shot pulses or a total energy $\gtrsim 10^{35} d_{\rm Gpc}^2$ erg. The energy
reservoir near the light cylinder limits the detection distance to $\lesssim
{\rm few} \times 100$ Mpc for a fluence $\sim 1$ Jy ms unless conditions are
more extreme than for the Crab pulsar. Similarly, extreme single pulses from
ordinary pulsars and magnetars could be detectable from throughout the Local
Group and perhaps farther. Contributions to dispersion measures from galaxy
clusters will be significant for some of the bursts. We discuss tests for the
signatures of bursts associated with extragalactic NSs.
\end{abstract}

\section{Introduction}
\label{sec:intro}
Over the last decade  individual radio bursts have been  found as a byproduct of surveys for
periodic radio pulsars.   Reobservations ultimately reveal that many of the sources are radio pulsars, differing only in the means used to initially discover them, but otherwise having similar millisecond  pulse widths  and dispersion measures (DMs) consistent with a Galactic origin.  However, a minority of the bursts has defied redetection.   Some are clearly Galactic
\citep[e.g.][Figure 3]{2006Natur.439..817M, 2014ApJ...790..101S}  but others have DMs much too large to be accounted for by the modeled foreground interstellar medium (ISM) in the Milky Way.    In the literature, bursts found through single-pulse detection algorithms have loosely been referred to as `rotating radio transients' \citep[RRATs;][]{2006Natur.439..817M}
and are consistent with a Galactic population of neutron stars \citep[NS;][]{2014ApJ...790..101S} .  The more recently discovered events with  DMs too large to be accounted for by the Galaxy have been termed `fast radio bursts' (FRBs)
\citep[][]{2007Sci...318..777L, 2011MNRAS.415.3065K, 2013Sci...341...53T, 2014ApJ...790..101S, 2014ApJ...792...19B, 2015ApJ...799L...5R, 2015MNRAS.454..457P} and appear to be extragalactic in origin.  To avoid confusion,
we use the term `extragalactic radio burst' (ERB) for these events.\footnote{For example, an FRB candidate event could  be Galactic if the large DM is due to electrons in a Galactic HII region \citep[e.g.][]{2014MNRAS.440..353B}, in which case the source might
be termed an RRAT.}   

We assume that most if not all of the reported ERBs are in fact astrophysical in nature. 
Some similar bursts are due to radio-frequency interference (RFI) \citep[`perytons;'][]{2011ApJ...727...18B, 2014ApJ...795...19S}  because they have time-frequency signatures inconsistent with  pulsar-like dispersion delays and they have  been  associated with  specific sources of RFI \citep[][]{2015arXiv150402165P}.   Though the number of  detected ERBs is small, the inferred rate is surprisingly large, $\sim 10^3$-$10^4$ bursts d$^{-1}$ 
\citep[][]{2015MNRAS.447.2852K, 2015ApJ...807...16L}.

In this paper we examine the requirements on radiation coherence and source demographics that are implied by ERB properties summarized in Section~\ref{sec:salient}  and analyze conditions under which they can be met by the population of extragalactic NS.   This requires a discussion  of coherent radiation bright enough to be detectable in a volume 
containing enough sources to account for the net observed rate of ERBs for plausible event rates per source.  Our paper therefore considers the seemingly disparate topics of
radiation physics and populations of compact objects.   Other interpretations of ERBs must also consider these issues, so  much of our discussion is generic even if NS are not the source population.
 
The first of these two major topics --- radiation --- is discussed in Sections~\ref{sec:maxobserved}-\ref{sec:cr}. 
In Section~\ref{sec:maxobserved} we summarize  giant pulses (GPs)  from the Crab pulsar, which serve as exemplars for discussing ERBs.  We demonstrate that there is more than enough free energy available to account for observed ERB flux densities emitted from at least 100~Mpc and we consider extension of the pulse-amplitude distribution for the Crab to very long waiting times between super-GPs that might correspond to ERBs. We also emphasize that  the largest individual shot pulse ever detected \citep{2007ApJ...670..693H} is comparable to those required for ERBs and therefore underlies why it is completely reasonable
on an empirical basis  to consider extragalactic NS as a possible source class for ERBs. 
Section~\ref{sec:amsn} discusses the shot-noise model  that incoherently combines coherent shot pulses in order to match   observed ERB properties.

Section~\ref{sec:cr} considers a specific radiation process, coherent curvature radiation,  to demonstrate detailed requirements in the context of a NS magnetosphere.  We test our results against the  largest Crab shot pulse and find that a large number of leptons ($\sim 10^{29}$) with a relatively small net charge ($\sim 10^{21}$ e) must radiate coherently to account for the measured amplitude; the small net charge reflects the importance of radiation reaction \citep[e.g.][]{1976MNRAS.177..109B}. The required charge density in the coherently emitting region is large, perhaps $\sim 10^{11}e\,{\rm cm^{-3}}$, 
using coherence volumes (estimated in Appendix~\ref{app:Coherence}), which are $\gtrsim r_c\lambda^2$ for coherent radiation from charges streaming along field lines with curvature 
radius $r_c$. Note that the coherence volume is $\sim (r_c/\lambda)\times\lambda^3\gg\lambda^3$,
i.e. far larger than is often assumed in discussions of coherent radiation \citep[e.g.][]{2000ApJ...535..365K}. Nevertheless, the required density is much larger than the Goldreich-Julian density. We do not discuss how these large density contrasts are produced, although we note a few instabilities that might be relevant.
Radiation reaction is likely a {\sl generic} issue for understanding ERBs regardless of the class of objects that cause them, including the collapse of supramassive NS \citep[][]{2014A&A...562A.137F}, evaporating black holes \citep{1977Natur.266..333R}, and cosmic strings \citep[][]{2014JCAP...11..040Y}, etc.

Section~\ref{sec:pops}  discusses the second major topic --- source population requirements ---  
where we estimate event rates per object needed from an  extragalactic NS population to  yield the large inferred  ERB rate.   We find that   a NS population extending to $z \sim 1$ is large enough that only a few large bursts per NS will suffice to account for the apparent ERB rate.  In this case, gravitational microlensing would  play a role.  However, we favor a closer population for which lensing is unimportant and has less stringent requirements on the radiation process, but requires 
 more bursts per object.    
 
 Section~\ref{sec:summary} summarizes how ERB source populations can be constrained and
 Section~\ref{sec:tests} discusses observational tests for features of the NS model for ERBs. 
Section~\ref{sec:conclusions}  summarizes our overall results and conclusions.
Appendix~\ref{app:flux} gives  the calculation of flux densities of 
single and incoherently-combined shot pulses. Appendix~\ref{app:Coherence} presents the calculation of coherence volumes in dipolar magnetospheres.

\section{Salient Features of ERBs}
\label{sec:salient}

ERBs have  fundamental properties that need to be accounted for in any 
interpretation.  
First, millisecond  duration pulses  with flux densities $\sim 1$~Jy  must originate from  compact sources with relativistic flows. The radiation  must have high spatial coherence to produce the 
 very high inferred brightness temperatures, 
 \be
T_b = 
	\frac{S_\nu}{2 k}  \left(\frac{d}{\nu\Delta t}\right)^2
	\sim 3\times10^{35}{\rm K}\, S_{\nu, \rm Jy} (d_{\rm Gpc} / \nu_{\rm GHz} \Delta t_{\rm ms})^2,
\ee
 where distances are in Gpc and the burst duration has been associated with the light-travel time across the source. 
ERBs therefore are not unlike  pulses from Galactic pulsars  and RRATs so they could have similar underlying radiation physics.  

 Second is that ERBs, like pulsar pulses, have radiation time scales that must span a range from
 $\lesssim 1$~ns to $\gtrsim 1$~ms.  Nanosecond structure necessarily follows from the GHz 
 frequencies at which ERBs have been detected while ERBs manifestly show total widths up to 
 a few milliseconds.   Shot noise models with variable amplitudes or rates  naturally account 
 for these features.

Third, while arguments have been made that the high  dispersion measures  of ERBs originate
from or are mimicked by an emission process in the sources themselves \citep[e.g.][]{2014ApJ...797...70K, 2014MNRAS.tmpL...2L},  the simplest interpretation at present is that 
the burst sources are extragalactic \citep[e.g.][]{2014ApJ...785L..26L, 2014MNRAS.443L..11D}. 
\citet[][]{2015arXiv150505535C} suggest that the non-Galactic portions of ERB DMs are dominated by plasma in supernova shells from relatively nearby (non-cosmological) galaxies. 
 Extragalactic distances require larger radio luminosities than typical radio pulsars, however, and one goal of this paper is to examine whether coherent processes in  neutron-star (NS) magnetospheres (or similar objects) can produce detectable bursts originating from large distances. 

 Finally, despite the small number found so far, the  implied aggregate event rate   $\sim 10^3$-$10^4$~day$^{-1}$ over the entire sky   is astonishingly large
\citep[][]{2015ApJ...807...16L, 2015MNRAS.447.2852K}, much  larger than the observed rate of  gamma-ray bursts,
for example \citep[][]{2013Sci...341...53T,2014ApJ...790..101S}. The lack of any new events  in reobservations of  ERB sky directions requires either that ERBs originate from non-repeating astrophysical cataclysms  or  from repetitive sources with  very low burst rates.   Until a distance scale is established for ERBs, it is not possible to identify any particular underlying source class.  However, in this paper we find that extragalactic NS could account for the burst properties observed so far.

\section{Maximal Radio Emission from Pulsars}
\label{sec:maxobserved}
In this section we summarize large-amplitude coherent radiation from pulsars as 
a benchmark for discussion of ERB radiation.   
Rotation-driven NS dissipate their energy through emission 
 in particles accelerated by rotationally-generated electric fields,  low-frequency waves, and high-energy radiation. 
It is well  known from early studies of pulsars that relativistic particles have significant 
energy losses from coherent  radio emission  \citep[][]{1970Natur.227..465S,  1971ApJ...164..529S, rs75, 1976MNRAS.177..109B}.
It  follows, therefore, that the amplitudes and durations of narrow pulses are limited by  radiation reaction, a point 
emphasized by \citet[][]{1980MNRAS.190..945B} in their analysis of  the Crab pulsar. 
But this also means that coherent radio emission can tap a sizable fraction of the free energy available in particles.  
Radio emission  is typically a tiny fraction
of the spindown energy loss rate,  $\Edot = I\Omega\dot\Omega$ ($\Omega$ = spin frequency and $I$ = moment of inertia), particularly for  the  average emission from objects with large values of $\Edot$, such as the Crab
pulsar.  However, some older, long-period pulsars have
$L_r /\Edot \sim$~1 -  10\% 
\citep[e.g.][]{2002ApJ...568..289A}.  Recent work has also revealed a direct link between radio emission and $\Edot$ in
intermittent pulsars  showing on-and-off intensity transitions that accompany large changes in $\Edot$
\citep[e.g.][]{2006Sci...312..549K}.   

\subsection{ Giant Pulses from the Crab Pulsar}
\label{sec:crab}

In this paper, we use GPs from the Crab pulsar as benchmarks in several ways.  First, we can extrapolate the amplitudes of GPs to those of young-pulsar analogs in other galaxies and consider whether they can serve as an underlying population for ERBs.  Second, ERBs may originate from older NS due to triggering by unknown agents of Crab GP type emission.  Third, the narrowest features seen on ns time scales in a few Crab GPs provide evidence for extremely bright emission from individual coherently radiating regions that can be superposed in large numbers to provide detectable pulses from large distances.  Thus, ERBs might be from young pulsars but they need not be. 

\begin{figure}
\begin{center}
\includegraphics[scale=0.6]{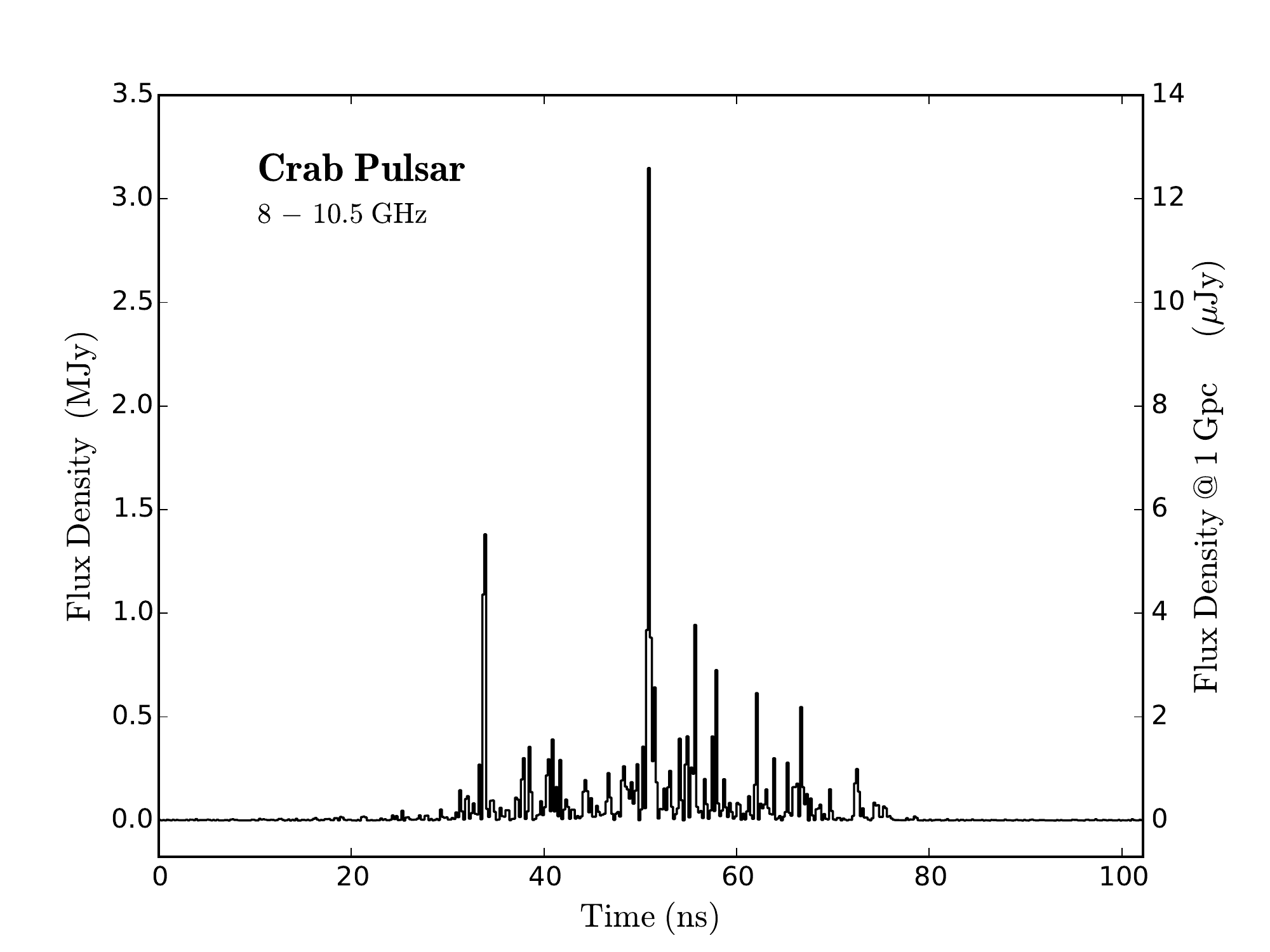}
\caption{
Portion of a single pulse from the Crab pulsar showing sub-ns structure (data courtesy of 
\citet[][]{2007ApJ...670..693H}) that is consistent with a shot-pulse model for the radiation.  The largest shot pulse exceeds 3~MJy and several others exceed 0.5~MJy.  The right-hand axis indicates the flux-density scale if the source were at a 1~Gpc distance; 
\label{fig:Megapulse} 
\label{fig:amsn}
}
\end{center}
\end{figure}

The brightest GP detected  so far had a peak amplitude $\Snumax  = 2.2$~MJy at 9 GHz 
and a pulse width $\Delta t$ smaller than the
$0.4$~ns time resolution  provided by the 2.5~GHz receiver bandwidth \citep[][]{2007ApJ...670..693H}. 
The implied brightness temperature    $T_b \gtrsim 10^{41.3}$~K
obviously requires  a coherent emission process. The upper bound on $\Delta t$ implies
a time-bandwidth product $\nu\Delta t\lesssim 4$,   suggesting that it is a single shot pulse produced by an individual, coherently emitting region. 
Figure~\ref{fig:Megapulse} shows this pulse with higher analysis resolution and a slightly different flux-density (T. Hankins and J. Eilek, private communication). 
Unlike this special case, most GPs, like all pulsar radiation,  have time-bandwidth products $\nu\Delta t \gg 1$  because they are {\sl incoherent superpositions of  shot pulses}.
Relatively frequent GPs at 0.43~GHz with  
$\sim 100$-200~kJy maxima  occur about once per hour
\citep[][]{1975MComP..14...55H, 2004ApJ...612..375C, 2010ApJ...722.1908C}.
They are broadened by interstellar scattering 
to  $\sim 100~\mu s$ 
from intrinsic widths of about $50~ \mu s$.  The implied time-bandwidth product
$\sim 50~\mu s \times 0.43~\rm GHz \sim 2\times 10^4$ { requires} a combination of coherent emission and
incoherent superposition and this is consistent with the appearance of single GPs obtained with high time resolution. The same can be said of pulsar signals generally.   
 
 Giant pulses from the Crab pulsar comprise $\sim 1$\% of all pulses and have an amplitude distribution  
that shows no discernible cutoff  in data sets as large as 
100~hr  of on-pulsar live time ($\sim 10^7$ pulse periods) 
\citep[][]{1995ApJ...453..433L, 2012ApJ...760...64M}.
Fits of power-law distributions $\propto \Snu^{-\alpha}$ yield
  exponents   ranging from $\alpha \approx 2.3 - 3.5$ in 
 observations at different frequencies (Mickaliger et al. 2013).
The power-law component may  be augmented  with a long tail associated with
additional, supergiant pulses (SGPs) 
\citep[][]{2004ApJ...612..375C, 2012ApJ...760...64M}, but with as-yet poorly constrained parameters 
for the tail.   The existence of  SGPs from the Crab pulsar suggests that much larger pulses from it and other objects may be seen only in  monitoring campaigns that are much longer than those conducted to date.

\subsection{GP Energetics}
\label{sec:energetics}

The largest GPs from the Crab pulsar  are easy to accommodate 
within the  total energy-loss budget of its magnetosphere but require  significant fractions of the available
particle energies to be directed into coherent emission. 

We  relate maximal GP emission to  the spindown energy loss rate 
$\Edot \sim 10^{38.7}$~erg~s$^{-1}$ by combining the 
radio emission efficiency $\epsilon_{\rm r} \le 1$
 with the beaming solid angle $\Omega_r$  into a combined
 factor $\zeta_r = (4\pi \epsilon_r/\Omega_r)$. The peak flux density is then 
\be
{S}_{\nu, \rm max} = 
	\left(\frac{4\pi \epsilon_{\rm r}}{\Omega_{\rm r}}\right)
	\left(\frac{ \Edot}{4\pi\Delta\nu_{\rm r}\dcrab^2}\right)	
\approx 
	\frac{ 100\zeta_r\, {\rm MJy}\,\,}{\Delta\nu_{\rm GHz}}
	\left(\frac{\Edot_{38}}{4.6} \right).
\ee
The approximate equality  uses parameter values for the Crab pulsar including  $\dcrab = 2$~kpc. 
Evaluating for the  2~MJy shot pulse at 9~GHz and the more frequent $\sim 100$~kJy pulses at 0.43~GHz,
we obtain $\zeta_r = 0.02$ and $0.002$, respectively. 
We may expect infrequent but very large pulses from the Crab pulsar when $\zeta_r \gg 1$, which may occur
as a consequence of favorable relativistic beaming or large instantaneous departures from the average
radio efficiency $\epsilon_r$.

The `Crab twin' pulsar B0540--69 in the Large Magellanic Cloud ($P = 50$~ms, $B = 5\times10^{12}$~G)
also shows giant pulses \citep{2003ApJ...590L..95J} but with   $\Snumax d_{B0540-69}^2$
smaller by a factor $\sim 0.1$ than those of the Crab pulsar.   Some of the difference could be due to less-favorable beaming for B0540--69.

While GPs can be accommodated by spindown losses, they may saturate the typical energy available in 
relativistic particles.     Force-free conditions
$\Evec\cdot\Bvec = 0$ in a neutron-star's magnetosphere require a charge (number) density
$\ngj \approx -\Omvec \cdot \Bvec / 2\pi e c$, the Goldreich-Julian (GJ) density,  
\be
{\ngj}_0 \approx  10^{10.84} \, {\rm cm^{-3}} B_{12} P^{-1}(R/r)^{3},
\label{eq:ngj0}
\ee
where the radial scaling applies to  the open-field-line region  of a dipolar magnetosphere with $r\ge R$, where $R$ 
is the NS radius,  and
$r\ll \rlc $ where the light-cylinder radius is $\rlc = cP/2\pi$.  For standard $\sim 10^{12}$~volt potential drops in polar cap models \citep[e.g.][Table V]{rs75, 1982RvMP...54....1M}, the particle energy loss rate 
 (assuming an electron mass) is approximately 
\be
\dot E_p = c A_{\rm pc} \ngj \gamma mc^2 
	\approx 10^{30}\,{\rm erg~s^{-1}} B_{12} R_6^3 \gamma_6 P^{-2}
	\propto \dot E^{1/2},
\ee
for a Lorentz factor  $\gamma$ and a  magnetic polar cap area $A_{\rm pc} \approx \pi R^3 / \rlc \propto P^{-1}$.   The particle loss rate scales as the square root of the spindown
loss rate in standard polar-cap models \citep[][]{rs75,1979ApJ...231..854A}.
The Crab pulsar ($B_{12} = 3.78$ and $P^{-1} = 30.2$~Hz) yields
$\dot E_p \simeq 10^{33.7}\,\rm erg~s^{-1}$.  
For the 9-GHz pulse described above, highly beamed radiation  with $\Omega_r/4\pi \lesssim 2\times 10^{-4}$ is required to provide  $L_r \lesssim \dot E_p$.  Since the beam solid angle  is likely even smaller than this limit for  
$\gamma \gtrsim 10^3$,   there is significant
headroom in the possible range of GP amplitudes to the extent that  particle flows and radiation
physics can allow much larger values. 

\subsection{Extrapolation of Rates and Amplitudes  from the Crab Pulsar}
\label{sec:gpextrap}

The maximum detectable distance of a Crab GP with amplitude $\Snu = 10^5$~Jy  is
$\dmax = \dcrab \left(\Snu/\Snumin\right)^{1/2} \approx 2.5$~Mpc  -- 
a distance that encompasses most local-group galaxies --  for   a detection threshold $\Snumin = 60$~mJy (e.g. a $5\sigma$ detection for an Arecibo-class telescope\footnote{Assuming a bandwidth of 50~MHz and  a system equivalent flux density of
4~Jy along with a matched filter with 1~ms width. }).   Flux densities about
$10^5$ times brighter ($\sim 10^{10}$~Jy)   are needed  for analogous objects to be detectable 1~Gpc away.

We scale from  Crab GPs to fiducial Gpc distances using 
\be
\Snumax = 
{S}_{\rm \nu, max, Crab}
 \left(\frac{\zeta_r}{{\zeta_r}_{\rm Crab}} \right)
\left(\frac{\dcrab}{d} \right)^2
\left(\frac{\Edot_{38}}{4.6}\right)
	\approx 0.2\,\mu{\rm Jy}\,\,		
	\frac{\Edot_{38}} {\Delta\nu_{\rm GHz} d_{\rm Gpc}^2}
	\left(\frac{ \zeta_r}{ {\zeta_r}_{\rm Crab}} \right).
\label{eq:fnu}
\ee

Pulses with $\Snu \sim 1$~Jy require a combination of larger $\zeta_r$ (i.e. larger radio efficiency or smaller beaming 
solid angle) with larger spindown loss rates $\Edot$ and distances smaller than   1~Gpc. 
To be detectable at 1~Gpc with a 60~mJy threshold, 
$(\zeta_r / \zeta_{r, \rm Crab})\Edot_{38} \gtrsim 3\times 10^5$.
Objects undoubtedly exist with spindown loss rates
$\Edot\propto \dot P P^{-3}  $ much larger than that of the Crab pulsar.  
This could include
the Crab pulsar itself when it was born with 
$P~\sim 20$~ms compared to its present-day 33~ms  period \citep[e.g.][]{1991tnsm.book.....M}.  Magnetars with 
100 times larger magnetic fields could have $\Edot$ up to $10^6$ times larger than the Crab pulsar's
loss rate if they are born with 10~ms periods.   However, their birthrate is small and they spend little time
at short periods if measured  spin down time scales at longer periods can be extrapolated to earlier times \citep[c.f.][Table 2]{2014ApJS..212....6O}.  Magnetars are therefore 
  less likely to be responsible for ERBs unless they  are relatively nearby. 

\subsection{Extreme-Value Statistics and Waiting Times}
\newcommand{\rmax}{r_{\rm u}}

We now consider the sampling of pulses from a single object
in terms of 
the probability density function (PDF) $f_1(S)$ and the
cumulative distribution function (CDF) $F_1(S)$ for their amplitudes, where   the  subscript  $\nu$ on $S$ is dropped to simplify the notation. 
Most pulsars do not show GPs and have PDFs that are typically consistent with log-normal distributions \citep[][]{2012MNRAS.423.1351B}. GP amplitude  PDFs, however, are heavy tailed with a power-law type shape for a sizable  range of pulse amplitudes.  By `heavy tailed', we mean that
the PDF extends to amplitudes much larger than the mean.  

\begin{center}
\begin{figure} 
\includegraphics[scale=0.45]{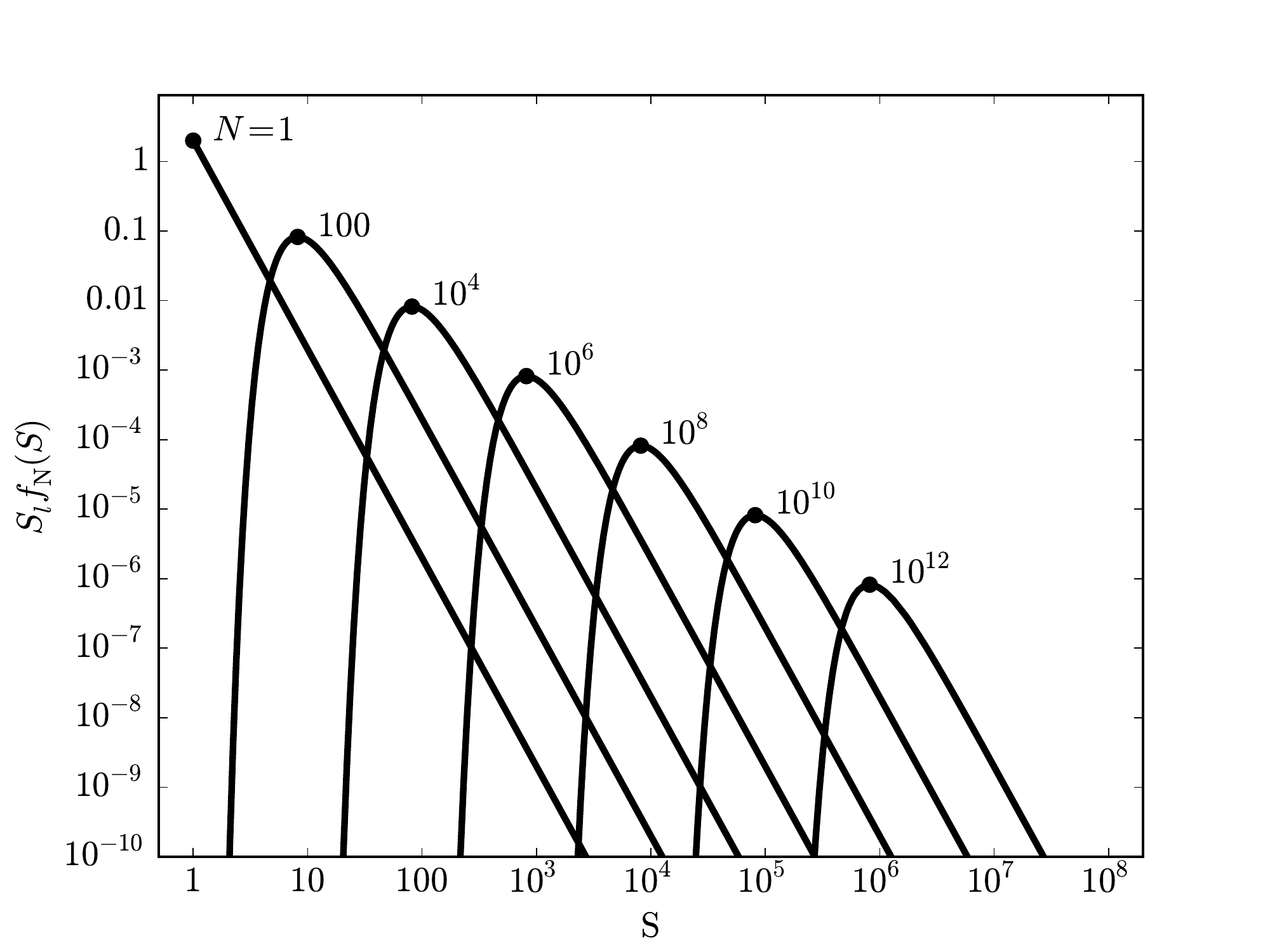}
\includegraphics[scale=0.45]{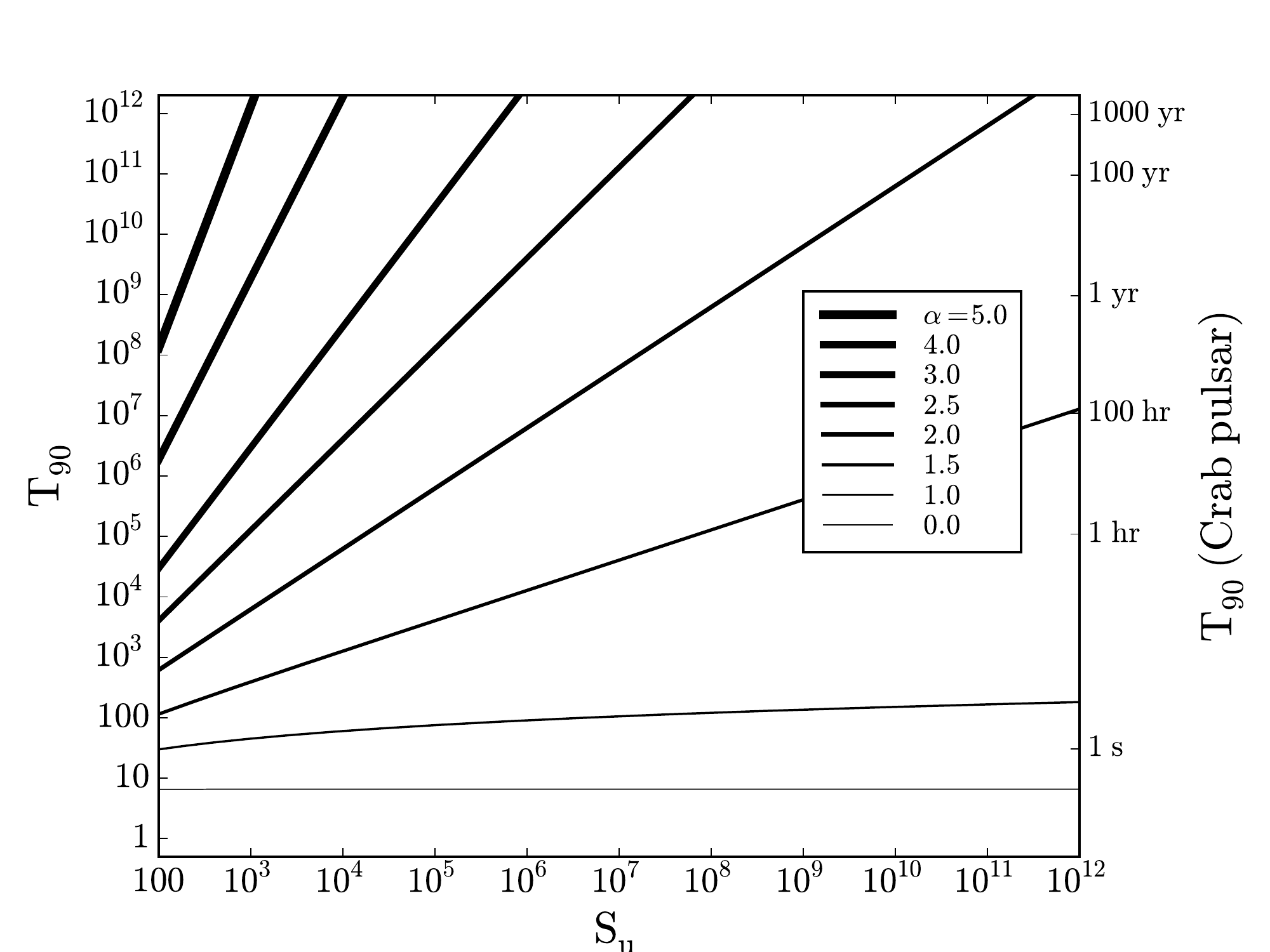}
\caption{
Left: PDF of maximum $S = r S_l$ for $\alpha=3$ and $r_u = S_l/S_u = 10^8$ for various $N$.
The filled circles denote the modes of the PDFs.
Right: Time $T_{90}$ (expressed in units of pulse period) to reach 90\% of the maximum amplitude $r_u$ for different values of $\alpha$. The right-hand scale gives $T_{90}$ in time units using
the spin period $P=0.033$~s of the Crab pulsar. 
\label{fig:pdf_of_max}
}
\end{figure}
\end{center}

We adopt  a  power law\footnote{A pure power-law model does not always appear to be a good model for histograms of pulse amplitudes from the Crab pulsar.   However, they always show long tails so we proceed by using power-law PDFs to make extrapolations while keeping in mind that the true form of the PDF may differ.} PDF, $f_1(S) \propto S^{-\alpha}$ between lower and upper cutoffs $S_l$ and $S_u$.
For steep power laws, an increasingly larger maximum $S$ is expected as the number of sampled pulses increases. 
Consider $N$ pulses  occurring  periodically in time $T=NP$ with period $P$  and amplitudes drawn from $f_1(S)$.  
The maximum amplitude has a CDF equal to the probability that all  amplitudes are less than or equal to $S$.   The extreme value CDF and PDF are therefore 
\be
F_N(S) = \left[ F_1(S) \right]^N 
\quad\quad \text{and} \quad\quad 
 f_N(S) = N f_1(S) \left[ F_1(S) \right]^{N-1}.
\ee 
Figure~\ref{fig:pdf_of_max} (left) shows a sequence of extreme-value PDFs for $\alpha = 3$ and
$S_u/S_l = 10^8$, which maximize at steadily larger flux densities as  $N$ increases from  $1$ to $10^{12}$.  Each  PDF is marked by its mode.  For $\alpha\ne 1$, $S_{\rm mode}  = S_l \left[1 - (N-1)(1-\alpha)/\alpha\right]^{1/(\alpha-1)} $.  The median is given by
$S_{\rm med} = \left [ 1 + e^{-(\ln 2)/N} \left(S_l/S_u \right)^{\alpha-1}\right]^{-1/(\alpha-1)}$.

A similar  approach can be applied  to pulses emitted  with Poisson statistics. 
The CDF is obtained by summing 
$F_N(S)$ over the Poisson probabilities for obtaining $N$ pulses given a mean number $\langle N \rangle$ in time $T$.   This approach would also be valid for the Crab pulsar because pulses
with amplitudes  above a fixed threshold occur at intervals consistent with  Poisson statistics \citep[][]{1995ApJ...453..433L}. The two approaches yield very similar results and in the following, we use  expressions for the periodic case. 

The time needed to see the largest pulses spanned by 
 the PDF $f_1(S)$ is strongly dependent on the upper cutoff $S_u$ and on $\alpha$. 
For any specified $S$, the  waiting time needed 
to see an amplitude as large as the  median  of $f_N(S)$  is obtained by solving $F_N(S) = 1/2$, 
\be
T = - P\ln 2 / \ln F_1(S).
\label{eq:Twait}
\ee
If we specify a median that is   within a factor
$f_u < 1$ of the largest possible amplitude $S_u$, the waiting time is
\be
T_f  = -P\ln 2 / \ln F_1(f_u S_u). 
\label{eq:Twait_f}
\ee

Figure~\ref{fig:pdf_of_max} (right) shows $T_f$ for $f_u = 0.9$ and multiple values of $\alpha$ that  demonstrate the reasonable result that steeper distributions (larger $\alpha$) with larger upper cutoffs $S_u $  require longer times to reach 90\% of the cutoff.   Flatter distributions, however, can be sampled in small amounts of time. 
The waiting times are $10^{12}$ periods  ($\sim  10^3$~yr) or larger for $\alpha \gtrsim 2$
and $S_u/S_l = 10^{12}$.   Since previous studies, as noted earlier, indicate $\alpha \sim 2.3$ to 3.5 for Crab-pulsar GPs,  the possibility remains that much larger pulses will occur if $S_u$ is large and observations extend over  time spans much longer than the 100~hr of previous studies.
Conversely, 
 continued monitoring of GPs from the Crab pulsar can determine or place  a lower bound on the upper cutoff $S_u$.  Knowledge of the cutoff can help establish the distance scale of  any  extragalactic GP-emitting objects, including  ERBs under the GP hypothesis,  or it could  rule out the hypothesis if $S_u$ is not substantially larger than GPs seen to date from the Crab pulsar.   The Bayesian posterior PDF for $S_u$ can be evaluated from an observation of the largest pulse $S_{\rm max}(T)$ seen in a time span $T$
 using assumed values for $S_l$ and $\alpha$ based on previous measurements.  For steep power laws, the resulting lower bound will be a modest multiple of  $S_{\rm max}(T)$. 
We remark here that in Section~\ref{sec:CrabShots} our analysis  of radiation physics  suggests that the largest possible giant pulses from the Crab pulsar may be constrained by the maximum fluence of individual coherent emitting structures.  If so, larger amplitude GPs would require greater numbers of coherent structures summing incoherently.

Now we compare the median amplitudes $S_{1,2}$ expected in two time intervals, $T_1 < T_2$.  From Eq.~\ref{eq:Twait}, 
\be
\frac{T_2}{T_1} 
= \frac{\ln F_{1}(S_1)}{\ln F_{1}(S_2)}
\approx \frac{1-F_1(S_1)}{1-F_2(S_2)}. 
\ee
Solving for $S_2$, we obtain
\be
S_2 \approx 
	S_1 \left(\frac{T_2}{T_1}\right)^{1/(\alpha-1)}
	\left[
	1 + 
	\left(S_1/S_u\right)^{\alpha-1}  \left(T_2/T_1 - 1\right)  
	\right]^{-1/(\alpha-1)}
	\approx 
	S_1 \left(\frac{T_2}{T_1}\right)^{1/(\alpha-1)},
\label{eq:r2max}
\ee
where the approximate equality holds for
$(T_2/T_1) (r_1/r_u)^{\alpha-1} \ll 1$.
Using the $S_1 \sim 100$ kJy pulse typically seen from the Crab pulsar in $T_1 =$~one hour (c.f. Section~\ref{sec:crab}), and using power-law indices that have been inferred
($\alpha \approx 2.3 - 3.5$) to extrapolate to long times,    
the largest pulse expected in  $10^3$~yr using Equation~\ref{eq:r2max}) is $S_2 = 10^{7.7}$~Jy ($\alpha=3.5$) to $10^{10.3}$~Jy ($\alpha=2.3$).   The corresponding maximum detection distances 
are    $\dmax = 15$~Mpc  to $300$~Mpc using
$\dmax = 0.63~\rm Mpc~(T/1~\rm hr)^{1/2(\alpha-1)}$.


\subsection{Implications for Weak and Strong ERBs}

We have shown in the previous section that a power-law PDF for GPs can lead to very long waiting times for the largest pulses if a steep PDF extends to very high values.     
Here we consider the expected number of weaker pulses  above   some threshold in the same time interval containing a very bright pulse.  
The number of detected pulses, defined as those exceeding a threshold $S$,  out of $N$ total
pulses  is, on average,$N_d(S)   = N \left [1 - F_1(S)\right]$. From this $N$ can be expressed in terms of $S_{\rm max}$ and $N_d$.   The brightest pulse $S_{\rm max}$
corresponds to $N_d(S_{\rm max}) = 1$, so the number above a lower threshold $S_{\rm th} < S_{\rm max}$
is 
\be
N_d(> S_{\rm th} \vert S_{\rm max}) = N \left [1 - F_1(S)\right]   
=
\left ( \frac{S_{\rm max}}{S_{\rm th}}\right)^{\alpha-1}
\left[ 
\frac{1-(S_{\rm th}/S_u)^{\alpha-1}}{1 - (S_{\rm max}/S_u)^{\alpha-1}}
\right]
\approx 
\left ( \frac{S_{\rm max}}{S_{\rm th}}\right)^{\alpha-1},
\label{eq:Nd}
\ee
where the approximate expression is for $S_{\rm th}, S_{\rm max} \ll S_u$. 
For a threshold a factor of ten lower than the maximum pulse, we expect
$N_d(> S_{\rm th} \vert S_{\rm max}) \approx 10^{\alpha-1}$ or 20 to 320 weaker pulses for
$\alpha = 2.3$ to $\alpha=3.5$, respectively.     

Extrapolation from the Crab pulsar  to an individual ERB source  suggests that a large number of weak pulses should be seen for the cases where the ERB detection is well above an overall detection threshold.   The four ERBs detected by 
\citet[][]{2013Sci...341...53T} are above threshold by factors of 1.2 to 5.4.   However, the dwell time on each sky position was only 270~s,  so the expectation of a larger number of weaker pulses depends heavily on assumptions about the overall {\it population} of ERB sources.   The Lorimer burst
\citep[][]{2007Sci...318..777L} exacerbates the situation because its amplitude is 100 times threshold, but the overall issues about the luminosity function and spatial distribution of ERB sources remain the same.  

If all sources have the same intrinsic amplitude PDF, weaker pulses should have been detected from all of the ERB sources at the same distance as, for example, that of the largest of the four ERBs.   Additionally, a uniform population of sources would produce a population  PDF $\propto S^{-5/2}$ in Euclidean space for standard-candle
luminosities or a broader distribution for the intrinsic power-law PDF considered above.  

That weaker pulses have not been detected suggests several alternatives.  One is that the amplitude distribution for ERBs  is dissimilar from the steep power law seen for most of the GPs from the Crab pulsar.  The Crab pulsar itself suggests that supergiant pulses (SGPs) occur as a long tail with a different distribution than `regular' GPs \citep[][]{2004ApJ...612..375C, 2012ApJ...760...64M}, as mentioned in Section~\ref{sec:crab}.   To account for the absence of weaker ERBs in Thornton et al.'s results, a suitable  intrinsic amplitude  PDF would have to be combined with a spatial population distribution that excludes the continuous, steep {\it net} amplitude PDF expected in Euclidean space for standard candles (or for any given luminosity out of a broad luminosity function).    One possibility is for the spatial distribution to be highly clustered so that there is a gap in the population amplitude PDF not far above the amplitudes of the ERBs detected so far.

Stronger pulses (than ERBs observed so far)  are not  an issue (yet) in spite of  the large number of ERB sources needed to account for the inferred all-sky rate.   Suppose the four ERBs detected by  
\citet[][]{2013Sci...341...53T} originate from the same distance $D_{\rm FRB}$ and from a standard candle distribution, while  ignoring the differences in amplitude (which are not large but are influenced by pulse-scattering broadening for one out of the four and by off-axis telescope gain differences). Any  nearer sources are fewer in number for a uniform source density,  so less than one, stronger ERB is expected for distances $D < 4^{-1/3}D_{\rm FRB}$ corresponding to amplitudes 
$S > 2.5 S_{\rm ERB}$.   A corresponding statement can be made for a non standard-candle
distribution.  Radiation physics may a provide a maximum possible luminosity that would cap the maximum observable amplitude.

We consider these issues further in the next three sections that consider ERB radiation physics and population distributions.

\section{Shot Noise Model for ERBs}
\label{sec:amsn}

The fundamental unit of radio emission is
 a coherent shot pulse with unit time-frequency product (a consequence of Fourier transforms), so detection at radio frequencies $\nu \gtrsim 1$~GHz requires  widths  $W_s \lesssim 1$~ns.  
 Actual pulses have large time-frequency products  $\nu W_i \gg 1$ because they are incoherent combinations of  coherent shot pulses.   
 In this section we discuss the requirements for generating shot pulses that are  sufficiently bright and numerous to account for ERB amplitudes after they are combined incoherently.

\subsection{ Maximal Shot Pulse Amplitudes}
\label{sec:maxshot}

First we calculate the maximum possible  radiation from a single coherently-emitting entity so that we can quantify the number of  shot pulses needed to account for the amplitudes of ERBs and then  assess plausible maximum distances of ERB sources.    ERB detections may in fact involve observational selection of extreme physical conditions.
An individual emitter  might be a charge clump  in the 
relativistic flow of a magnetosphere 
\cite[][]{1991ApJ...378..543W, 1998ApJ...506..341W, 2003Natur.422..141H, 2006MNRAS.369.1469A}.
Clumps may be produced from a two-stream instability or  a bunching
instability associated with coherent synchrotron radiation
\citep[][]{1971ApJ...170..463G, 1978MNRAS.185..493B, 2005PhRvE..71d6502S,2006PhRvS...9k4401S, 2005PhRvE..72b6410S}.

A simple estimate for { maximum possible} shot-pulse amplitudes is as follows.  Consider a dense, relativistic charge clump    that radiates a fraction $\epsilon_E$ of its  total energy $E_N = N\gamma mc^2$ ($N$ = number of particles)  into a solid angle $\Delta\Omega \sim \gamma^{-2}$ in a time $W_s$ as seen by 
an observer.  The  unit time-bandwidth product of the pulse implies that  the spectrum extends to an upper frequency $\nu_s \sim 1/W_s$ with a spectral shape $\snu$ normalized to unit area.  For a receiver bandwidth $\Delta\nu_r \ll \nu$ at a centre frequency $\nu\lesssim \nu_s$
and a source distance $d = 1~{\rm Gpc}~d_{\rm Gpc}$, the peak flux density of the shot pulse\footnote{Here and in Appendix~\ref{app:flux} we use lower-case $s_\nu$ to designate the flux density of a shot pulse and upper case $S_\nu$ for the macroscopic ERB pulse comprising an ensemble of shot pulses.} is 
(c.f. Appendix~\ref{app:flux})
\be
\sshotmax = \frac{\epsilon_E E_N  \Br f(\nu)}{\Delta\Omega d^2}   	
          = \frac{\epsilon_E \gamma^3 N mc^2 \Br f(\nu)}{(\gamma^2\Omega) d^2} 
          \approx 10^{-29} \epsilon_E N \gamma_3^3 d_{\rm Gpc}^{-2} (\gamma^2\Omega)^{-1} {\Br}_{\rm GHz} \snu  \, \rm Jy,    
	\label{eq:Snu}
\ee
where the electron mass has been used to evaluate the expression
 and the factor $\gamma^2\Omega$ indicates that the relevant
solid angle may differ from the assumed $\Omega = \gamma^{-2}$. 
A 1-Jy shot pulse at $d = 1$~Gpc requires the energy in {\it at least} 
$N=10^{29} \epsilon_E^{-1}$ 
electrons and/or positrons
 for nominal values of other parameters in this estimate.

The required number of particles  can be expressed
as $N = \etagj\eta_V {\ngj}_0 \lambda^3$ where $\etagj$ is a multiplier 
of the surface GJ density ${\ngj}_0$ (Equation~\ref{eq:ngj0}) and  $\Vc = \eta_V \lambda^3$ is the `coherence' volume from which particles produce a coherent pulse  measured by an observer.  
It is often assumed that $\Vc \sim \lambda^3$  or $\eta_V \sim 1$  
\citep[e.g.][]{1977MNRAS.179..189B, 1999ApJ...517..460S, 2000ApJ...535..365K}, but  for relativistic motion in 
a magnetic dipole,  we find 
$\eta_V \gg 1$ (Appendix~\ref{app:Coherence}).
To provide the number of particles, the combined multiplier must satisfy 
\be
\etagj \eta_V \approx 10^{14} 						
		\frac{P  \nu_{\rm GHz}^3}
		       {\eps_E \gamma_3^3 \Br \snu B_{12}}
		       \left( \frac{\sshotmax d^2}{1~\rm Jy ~ Gpc^2} \right),
\ee
where we have substituted for the surface GJ density from Eq.~\ref{eq:ngj0}.
The required multiplicity factor can be  reduced significantly under several conditions:    
shot pulses  that have peak flux densities $\sshotmax \ll  1$~Jy; small spin periods,
large  magnetic fields and Lorentz factors,
and distances less than 1~Gpc.   

As calibration of our analysis, we revisit  the 2~MJy shot pulse  from the Crab pulsar at 9~GHz, which  becomes 2~$\mu$Jy  if   emitted at 1~Gpc,  or 
$\sshotmax d^2 = 8~ \rm MJy~ kpc^2 = 8\times 10^{-6}~\rm Jy~Gpc^2$.  
This shot pulse itself requires a large multiplier,   
\be
(\etagj \eta_V)_{\rm Crab}  
		\approx 
		\frac{10^{9.7}d_{\rm Gpc}^2 }
		       {\eps_E \gamma_3^3 \Br}.
\ee
Pair production  accounts for some of the multiplicity, typically $10^3$ to $10^4$ in pulsar models \citep[e.g.][]{rs75},  and an $e^{\pm}$ Lorentz factor $>10^3$ could decrease the required value.  The remaining multiplicity can come from the volume factor.  Emission far from the NS surface increases the required multiplicity  because the particle density scales as $r^{-3}$, so high altitude emission regions require a volume multiplier  comparable to that needed for ERBs.

  
\subsection{Incoherent Summation of Shot Pulses}
\label{sec:incohsum}

 An empirical description of pulses from both pulsars and ERBs is the
amplitude modulated noise model   \citep[AMN;][]{1975ApJ...197..185R}  in which
a Gaussian  noise process with characteristic time scales $\lesssim \nu^{-1} \lesssim 1$~ns   is modulated on  time scales of microseconds and longer.  While the AMN model
is an adequate statistical description of  macroscopic pulses, a better physical description is 
amplitude and rate modulated shot noise,   a sequence  of shot pulses with a time-dependent Poisson rate
\citep[][]{1976ApJ...210..780C, 2004ApJ...612..375C, 2011MNRAS.418.1258O, 2011ApJ...733...52G}.      
In Appendix~\ref{app:ramsn} we demonstrate that amplitude and rate modulations are indistinguishable unless individual shot pulses can be identified in high-resolution time series.  
ERBs are too weak to allow detection of individual shot pulses and the extragalactic hypothesis implies that sources at Gpc distances will require the rate of shot pulses to be large, much larger than shown in Figure~\ref{fig:amsn}. 

Eq. ~\ref{eq:Snuapprox} gives the peak   flux density $\Snumax$ for an ERB lasting $W_i$ (e.g. $ \sim 1$~ms) in terms of the number of incoherently-combined  shot pulses $N_i$ of
amplitude $\sshotmax$. For time resolution $W_r = \Br^{-1}$ with a receiver bandwidth $\Br$,
\be
\Snumax \approx   (W_r / W_i)\, N_i\sshotmax.
\ee 
 The peak flux density is smaller than the flux density summed over all shot pulses $\sim N_i\sshotmax$ by the factor $W_r / W_i \ll 1$.   Smoothing of the intensity does not change this result (Appendix~\ref{app:ramsn}) because the frequency range of the receiver is the determining factor.   
 For example, $W_i = 1$~ms and  $W_r = 10$~ns for a receiver with a 100~MHz bandwidth, yielding $W_r / W_i \sim 10^{-5}$.   If $N_i$ is very large, individual shot pulse amplitudes $\sshotmax$ can be much smaller than the measured peak flux density $\Snumax$.

 To investigate what is required,
let the ERB arise from a volume 
$\Vi\sim (cW_i)^3$ containing $N_i$ coherently-emitting regions that produce  incoherently combined shot pulses. Each    has a subvolume
$\Vc$ so
the ensemble of  $N_i$ regions  fills a volume 
$N_i \Vc \equiv \zeta_{\rm ff}\Vi$, 
where $\zeta_{\rm ff}<1$ is a volume filling factor, or 
$N_i=\zeta_{\rm ff}\Vi/\Vc$. 
Eq. (\ref{volume}) indicates that the characteristic size of the coherent volume is
$\Vc/(cW_s)^3\lesssim\omega r_c/c\sim\gamma^3$ if the coherent radiation at frequency $\nu=\omega/2\pi$ is
from particles streaming along magnetic field lines with radius of curvature $r_c$; actual values are model dependent and may be larger or smaller than this value.
The typical peak shot-pulse  amplitude is then
\be
\sshotmax 
	\approx
	\frac{\Snumax}{N_i} 
	\frac{W_i}{W_r}
	= 
	\frac{\Snumax}{ \zeta_{\rm ff}}
	\frac{\Vc}{\Vi}
	\frac{W_i}{W_r}
	\lesssim
	\frac{\Snumax}{ \zeta_{\rm ff}}
	\frac{\gamma^3 W_s^3}{W_i^3}
	\frac{W_i}{W_r} 
	= 
	\frac{\Snumax}{ \zeta_{\rm ff}}
	\frac{(\gamma W_s)^3}{W_i^2 W_r}
	\sim
	0.1
	\Snumax 
	\frac{{\Br}_{\rm GHz} (\gamma_3 W_{s, \rm ns})^3}{\zeta_{\rm ff} W_{i, ms}^2}	.
\ee
For nominal values in the rightmost equality, shot pulses are 10\% of the ERB amplitude.  However this is an upper bound and  the shot pulse width may be substantially shorter than 1~ns.  Observed widths are somewhat larger than 1~ms, although scattering in the ISM or intergalactic medium (IGM) may account for some of the observed widths.  Shot pulses therefore could be much smaller than the ERB amplitude by six orders of magnitude or more if the filling factor is not small.   This means that 
1~Jy ERBs could be accounted for by  shot pulses that are similar in pseudo-luminosity to the 
MJy shot pulse from the Crab pulsar.

\section{Coherent Curvature Radiation}
\label{sec:cr}

Using general arguments we have shown in Section~\ref{sec:amsn} that the largest shot pulse seen from the Crab pulsar requires
extreme radiation efficiency and particle concentration; even more extreme conditions are required to produce
$\sim$ Jy shot pulses from distances $\sim$ Gpc.  However, incoherent summation of shot pulses, as noted, can allow the required shot pulse amplitudes to be comparable to the maximum seen so far from the Crab pulsar.  Here we derive the requirements for the specific case of 
curvature radiation.

   \subsection{Fractional  Clump Charge}
  
A simple estimate of the net fractional charge that can produce a consistent result is as follows.  
A single electron moving on a curved path with 
radius $r_c = 10^9 r_{c,9}$~cm radiates power $P_1 = 2\gamma^4 e^2 c/3 r_c^2$ 
\citep[e.g.][]{1962clel.book.....J}
and has a long radiation lifetime, 
\be
\tau_1 = \frac{\gamma m_ec^2}{P_1} = \frac{3}{2} \frac{r_c^2}{\gamma^3 r_e c} = 10^{11.2}~s\, \gamma_3^{-3} r_{c,9}^2.
 \ee
 The lifetime of
  $N$ charges of the same sign radiating  coherently at a rate $P_N = N^2 P_1$ is
 $\tau_N = \tau_1 / N$.   A  shot pulse with observed duration $r_c / \gamma^3 c$
 corresponds to a time interval  $r_c / \gamma c$ over  which the clump radiates toward the observer. 
 However, if radiation losses cause the clump lifetime to be   shorter than this time, $\tau_N < r_c / \gamma c$,  
 the observed pulse width will be narrower, 
 $\tau_N/\gamma^2$,   and the spectrum correspondingly wider.   
   
For the nominal particle number $N\sim 10^{29}\epsilon_E^{-1}$ calculated earlier, the lifetime 
$\tau_N = 10^{-17.8}\epsilon_E^{2}$~s is far too short to yield enough power in the radio band.  
The solution is  for the clump
of  $N$ particles to be nearly charge neutral.  Letting  the effective charge 
be $Q_c e$, the  clump lifetime becomes  $\tau_c = N \tau_1 /Q_c^2$ and matching  it to 
$r_c / \epsilon_E\gamma c$ yields
\be
\frac{Q_c}{N} = \left(\frac{\epsilon_E\gamma c \tau_1}{N r_c}   \right)^{1/2}
	=  \left(\frac{3 r_c\epsilon_E}{2r_e \gamma^2  N}   \right)^{1/2}
	= 10^{-6.7}\epsilon_E \gamma_3^{-1} r_{c,9}^{1/2}(\epsilon_E  N)_{29}^{-1/2}
	= 10^{-6.7}\epsilon_E \left[\frac{r_{c,9} \gamma_3 \Br \snu}{S_\nu d_{\rm Gpc}^2} \right]^{1/2}.
\ee  
The nearly charge-neutral  clump   can radiate 
 the  total energy available in the clump but at a rate that maximizes emission in the radio band.  We do not specify the mechanism that forms and holds together the clump, but it must radiate as a coherent unit  long enough to produce the    shot pulses with durations $\lesssim 1$~ns
 in the observer's frame.

\subsection{Coherent Shot Pulses from Curvature Radiation}
\label{cohshot}

We consider radiation from a relativistic clump having net charge $Q_e$ and  Lorentz factor $\gamma \gg 1$ that comprises
  $N$ particles, presumed to be electrons and positrons,  produced in some short-lived, burst-like magnetospheric event. 
The clump is presumed to move along a curved path (probably a particular
magnetic field line), emitting curvature radiation that is optimally beamed toward the observer
at some point along the path where the curvature is $r_c$. 
For notational simplicity in this section we use  $\omega=2\pi\nu$.
The observed flux density  is
\be
S_c(\omega)\approx\frac{3^{1/3}(\omega r_c/c)^{2/3}e^2q_c^2(\omega)\Dnud f(\xi)}{4\pi^2d^2c}
\label{fluxdensity}
\ee
where, for an observer situated
at an angle $\eps$ relative to the instantaneous plane of motion,
$\xi=(\omega r_c/3c)(2\kappa_0)^{3/2}=(\omega r_c/3c)(1/\gamma^2+\eps^2)^{3/2}$ and, if $\zeta=\omega r_c/3c\gamma^2$,
\be
f(\xi)=\xi^{4/3}\left[K_{2/3}^2(\xi)+\left(1-\frac{\zeta^{2/3}}{\xi^{2/3}}\right)K_{1/3}^2(\xi)
\right]
\label{synchform}
\ee
\citep{1962clel.book.....J};
the effective squared charge of the clump is
\be
q_c^2(\omega)=N+Q_c^2\Fco
\label{qsqeff}
\ee
where the first term arises from incoherent radiation and the second from coherent radiation.
In  the language of Eq.~\ref{eq:Snu},  the total energy radiated
during the portion of the orbit where emission is directed toward the observer is $\sim \gamma^4e^2q_c^2(\omega)/r_c^2c
\times r_c/\gamma c\sim \gamma^3e^2q_c^2(\omega)/r_c$ and
$f(\nu)\to f(\xi)/(\gamma^3c/r_c)$
is the frequency spectrum; $\Delta\Omega\sim(c/\omega r_c)^{2/3}$ at frequency $\omega\lesssim \gamma^3c/r_c$ 
and the fraction of the clump's energy radiated toward the observer is
\be
\epsilon_E\sim\frac{\gamma^2r_eq_c^2(\omega)}{Nr_c}
\approx \frac{5.9\times 10^{-17}\gamma_3^2q_c^2}{NP(r_c/\rlc)}~.
\label{epsE}
\ee

The form factor $\Fco$ is expected to be $\sim 1$ for coherent emission; this will be true
for observed frequencies $\omega\lesssim\omco$.
Conditions for coherence are reviewed in Appendix \ref{app:Coherence}.
For example, for a burst that lasts a time $t_b$ and 
emits charges along a single field line uniformly in time over timespan $t_b$,
$\Fco=[\sin(\omega/\omco)/(\omega/\omco)]^2$, with $\omco=1/2t_b$. In Eq. (\ref{fluxdensity})
we assume that the detector bandwidth $\Dnud$ is narrow compared with the relatively broad
band spectrum of instantaneous curvature radiation.

Eq. (\ref{fluxdensity}) implies pitifully small flux density unless coherence is substantial.
Thus, we presume that $\Fco\simeq 1$ for individual relativistic clump.
We assume that $\omco$ is below, but
not necessarily far below, the peak frequency $\ompk=2\pi\nupk=c\gamma^3/r_c$, where $r_c$ 
is the local curvature of the charged clump path when it is beamed toward the observer:
\be
\frac{\omco}{\ompk}=\frac{\nuco}{\nupk}=\left(\frac{\gamco}{\gamma}\right)^3
~~~~~\gamco=\left(\frac{\omco r_c}{c}\right)^{1/3}
=10^3(P\nucounit)^{1/3}(r_c/\rlc)^{1/3},
\label{nucodef}
\ee
where $P$ is the pulsar period in seconds, and $\rlc=c/\Omega=cP/2\pi=4.77\times 10^9P\,{\rm cm}$ is
the light cylinder radius. For comparison, the Lorentz factor required for the energy emitted in
{\sl incoherent} curvature radiation to be significant during a timescale $\dte=\epse r_c/c$
(defined as a fraction $\epse$ of the characteristic time $r_c/c$)  is
\be
\gaminco=\left(\frac{3r_c}{2r_e\epse}\right)^{1/3}=\frac{2.94\times 10^7P^{1/3}(r_c/\rlc)^{1/3}}{\epse^{1/3}}
=\frac{2.94\times 10^4\gamco}{(\epse\nucounit)^{1/3}}~.
\label{gamincodef}
\ee
The characteristic Lorentz factor associated
with pulsar voltage drops is
\be
\gamma_p=\frac{e\Omega^2\mu}{2m_ec^2}=\frac{1.29\times 10^7\mu_{30}}{P^2},
\label{gammap}
\ee
where $\mu=10^{30}\mu_{30}\,{\rm G\,cm^3}$ is the pulsar magnetic moment \citep[e.g.][]{rs75}.

To get a rough upper bound on the coherent emission from a single blob, we impose two constraints: (1) coherent
radiation dominates over incoherent radiation from the blob\footnote{Incoherent radiation from particles in the blob is completely different from the incoherent summation of  coherent shot pulses discussed in the paper.}  and (2) radiation losses are relatively minor
during emission. To keep the treatment simple, suppose that $r_c$ is roughly uniform along 
the particle path. 
Then the total energy radiated breaks
into two pieces: the coherent part, which is from $\omega\lesssim\omco$, where $\Fco\sim 1$, 
and the incoherent part that arises from $\omega\sim\ompk$
$\omega$:
\be
E_{incoh}\sim\frac{\epse Ne^2\gamma^4}{r_c}
~~~~~~~~~~~~~~
E_{coh}\sim\frac{\epse Q_c^2e^2(\omco r_c/c)^{4/3}}{r_c}\sim\frac{Q_c^2E_{incoh}}{N(\ompk/\omco)^{4/3}}
~.
\label{Erad}
\ee
We assume that $\dte\gtrsim r_c(2\kappa_0)^{1/2}/c\sim (r_c/c)^{2/3}\omega^{-1/3}$, 
which is the characteristic timescale {\sl at the emitter} for radiation toward
a particular observer at frequency $\omega$, but allow for the possibility $\dte\ll r_c/c$, 
in which case observation of the clump is highly fortuitous; consequently, with $\nu=\omega/2\pi=10^9\nuunit$ GHz,
\be
1\gtrsim\epse\gtrsim\left(\frac{c}{\omega r_c}\right)^{1/3}=\frac{10^{-3}}{
\nuunit^{1/3}P^{1/3}(r_c/\rlc)^{1/3}}~.
\label{epserange}
\ee
Near the lower bound, from the point of view of the observer
the clump  exists only for the time $\sim\omega^{-1}$ during which it emits
coherently toward the observer, a doubly lucky circumstance.
However, the very brightest shot pulses
are rare events that may involve unusual coincidences that correspond to observational selection.
Eq. (\ref{Erad}) shows that the coherent contribution dominates if
\be
\frac{Q_c^2}{N}\gtrsim\left(\frac{\ompk}{\omco}\right)^{4/3}~.
\label{cohdom}
\ee
Since the total energy contained in the burst is
$E_b=N\gamma m_ec^2$, then if we assume that coherent radiation dominates, radiation reaction 
does not severely limit the lifetime of a coherently emitting clump as long as
\be
\frac{Q_c^2}{N}\lesssim\frac{\gamma m_ec^2r_c}{\epse e^2(\omco r_c/c)^{4/3}}=\frac{c}{\epse\omco r_e}\left(\frac{\ompk}
{\omco}\right)^{1/3}=\frac{1.69\times 10^{13}(\ompk/\omco)^{1/3}}{\epse\nucounit}
\label{Qlimit}
\ee
Eqs. (\ref{cohdom}) and (\ref{Qlimit}) are consistent with one another provided that 
$\gamma\lesssim\gaminco$. 
Using Eq. (\ref{Qlimit}) in Eq. (\ref{epsE}) implies
\be
\epsilon_E\lesssim\frac{\ompk\omega^{1/3}}{[\epse(\omega r_c/c)^{1/3}]\omco^{4/3}}\approx
\frac{10^{-3}\ompk/\omco}{\epse(\nucounit P)^{1/3}}
\ee
which is $\sim 1$ for $\epse$ near its lower bound according to Eq. (\ref{epserange}).

\subsection{Application to the Crab Giant Shot Pulse}
\label{sec:CrabShots}

In order to understand the physical requirements for this model, we investigate the conditions necessary for
Eqs. (\ref{fluxdensity}) and (\ref{Qlimit}) to account for the brightest giant pulse from the Crab pulsar.
Eq. (\ref{fluxdensity})
requires a total charge $Q_c=10^{21}Q_{c,21}$, where
\be
Q_{c,21}=\frac{0.58\sqrt{S_c(\omega)d^2\Delta\nu_r^{-1}/{\rm MJy\,kpc^2\,GHz^{-1}}}}{\nucounit^{1/3}P^{1/3}(r_c/\rlc)^{1/3}
\sqrt{C(\omega/\omco)f(\xi)}}
\label{Qclump}
\ee
for coherent curvature radiation, where 
$\Gco=(\omega/\omco)^{2/3}\Fco$;
Eq. (\ref{Qlimit}) implies
\be
N\gtrsim\frac{2\times 10^{28}\epse\nucounit^{1/3}[S_c(\omega)d^2\Delta\nu_r^{-1}/{\rm MJy\,kpc^2\,GHz^{-1}}]}
{P^{2/3}(r_c/\rlc)^{2/3}(\ompk/\omco)^{1/3}C(\omega/\omco)f(\xi)}~,
\label{Nclump}
\ee
in agreement with the estimate in \S\ref{cohshot} if $\epse$ is near its lower bound in Eq. (\ref{epserange}),
and the total energy is
\be
E
\gtrsim\frac{1.63\times 10^{25}{\,\rm ergs\,}\epse\nucounit^{2/3}
[S_c(\omega)d^2\Delta\nu_r^{-1}/{\rm MJy\,kpc^2\,GHz^{-1}}]}
{P^{1/3}(r_c/\rlc)^{1/3}C(\omega/\omco)f(\xi)}~.
\label{Eclump}
\ee
The charge to mass ratio in the clump is therefore
\be
\frac{Q_c}{N}\lesssim\frac{2.9\times 10^{-8}P^{1/3}(r_c/\rlc)^{1/3}(\ompk/\omco)^{1/3}\sqrt{C(\omega/\omco)f(\xi)}}
{\epse\nucounit^{2/3}\sqrt{S_c(\omega)d^2\Delta\nu_r^{-1}/{\rm MJy\,kpc^2\,GHz^{-1}}}}~;
\label{chargetomasscrab}
\ee
thus, the clump must be nearly neutral, with a concentrated charge $\gtrsim 10^{21}$ to account for the brightest
Crab shot pulse.

The charge per lepton in a nearly neutral clump is $Q_c/N$, so we expect acceleration by an electric field
to result in $\gamma\simeq(Q_c/N)\gamma_p$, or
\be
\frac{Q_c}{N}=\frac{\gamco}{\gamma_p}\left(\frac{\ompk}{\omco}\right)^{1/3}=
7.75\times 10^{-5}P^{7/3} \mu_{30}^{-1}
\nucounit^{1/3}(r_c/\rlc)^{1/3}(\ompk/\omco)^{1/3}
\label{chargetomass}
\ee
consistency with Eq. (\ref{chargetomasscrab}) implies
\be
\frac{\epse}{(c/\omega r_c)^{1/3}}\lesssim\frac{0.377\mu_{30}\sqrt{C(\omega/\omco)f(\xi)}}{P^{7/3}(r_c/\rlc)^{1/3}
\nu^{1/3}\nucounit
\sqrt{S_c(\omega)d^2\Delta\nu_r^{-1}/{\rm MJy\,kpc^2\,GHz^{-1}}}}
\label{epslim}
\ee
Eq. (\ref{epslim}) favors fortuitously short lives for the brightest bursts,  $\epse\sim(c/\omega r_c)^{1/3}$,
and short spin periods.

To put the charge requirement in perspective, consider a coherent region
of volume $\mathscr{V}_s$ at radius $r$ with density contrast $C$ relative to the local Goldreich-Julian
charge density. Then the total charge is
\be
Q_c=\frac{\Omega\mu C\mathscr{V}_s}{2\pi ecr^3}\sim\frac{C\mathscr{V}_s}{2\pi er^3}\sqrt{\frac{cI\dot P}{P}}
=10^{31}I_{45}^{1/2}(\dot P/P)_3^{1/2}(C\mathscr{V}_s/r^3)
\label{Qcr}
\ee
where $I=10^{45}I_{45}\,{\rm g\,cm^2}$ is the stellar moment of inertia and $\dot P/P=(\dot P/P)_3/10^3\,{\rm y}$.
For the Crab's largest giant (shot) pulse $C\mathscr{V}_s/r^3\sim 10^{-10}$, or $(C\mathscr{V}_s)^{1/3}
\sim 15(r/10\,{\rm km})$ light-nanoseconds 
$= 2.4(r/\rlc)$ light-microseconds, which are both larger than $c/\nu\sim$ light-nanoseconds. 
Appendix \ref{app:Coherence} shows that coherently emitting volumes can be far larger than $(c/\nu)^3$. 
Eq. (\ref{volume}) implies volumes $\mathscr{V}_s\sim\gamma^3(c/\nu)^3$, or $\mathscr{V}_s^{1/3}
\sim\gamma c/\nu=\gamma_3/\nuunit$ light-microseconds, which is close to what is necessary for the Crab's largest
giant (shot) pulse if it originated near the light cylinder.

The energy requirement, Eq. (\ref{Eclump}), is impossible to assess without a detailed model for the formation
and acceleration of a clump. A standard of comparison for the clump energy in Eq. (\ref{Eclump}) is the total
magnetic energy in the unperturbed volume $C\mathscr{V}_s$ near radius $r$ i.e.
\be
E_B(r)=\frac{\mu^2C\mathscr{V}_s}{8\pi r^6}\sim\frac{I(\dot P/P)C\mathscr{V}_s}{4P(r/\rlc)^3r^3}
=\frac{7.9\times 10^{33}\,{\rm ergs}\,(\dot P/P)_3(C\mathscr{V}_s/r^3)}{P(\Omega r/c)^3}~.
\label{EBr}
\ee
For the Crab, Eq. (\ref{EBr}) implies $E_B(\rlc)\sim 2\times 10^{25}$ ergs for $(C\mathscr{V}_s/r^3)\sim 10^{-10}$,
which is comparable to Eq. (\ref{Eclump}), particularly if $\epse$ is near the lower bound implied by Eq. (\ref{epserange}).

This concordance suggests that the flux density of the brightest Crab pulsar giant pulse is
near the maximum that can be produced near the pulsar's light cylinder. Eq. (\ref{EBr}) 
could permit flux densities larger than $\sim$ MJy if the coherent emission arises somewhat
(but not very far) inside the light cylinder: Eq. (\ref{Eclump}) implies an energy bound
proportional to the flux density, whereas Eq. (\ref{EBr}) suggests an energy reservoir
$\propto r^{-6}$ at a fixed value of $C\mathscr{V}_c/\rlc^3$.
In view of the statistical considerations in Section 3.3, this concordance suggests that the high $S$ tail
   of the PDF of giant pulse flux densities for the Crab pulsar is relatively shallow.

\subsection{Application to Bursts from Extragalactic Neutron Stars}
\label{sec:xgal}

\def\Lnu{{\mathscr{L}}_\nu}
\def\Vi{{\mathscr{V}_i}}
\def\Snuth{S_{\nu,{\rm th}}}

A 1 Jy burst lasting 1 ms from a source at 1 Gpc distance corresponds to a luminosity
$\Lnu=1\,{\rm Jy\,Gpc^2\,ms}=10^{12}\,{\rm MJy\,kpc^2\,GHz^{-1}}$ in the observational frequency band.
In the curvature radiation model, $\Lnu$ determines the total charge squared, $Q_b^2$, associated
with the burst; $Q_b^2=N_i\langle Q_c^2\rangle$ for a burst consisting of $N_i$ incoherently-summed
shot pulses with mean square charge $\langle Q_c^2\rangle$ per coherently-emitting shot pulse.
Scaling from Eqs. (\ref{Qclump}) and (\ref{Eclump}) we find
\ba
Q_b&=&\frac{5.8\times 10^{26}\sqrt{\Lnu/{\rm Jy\,Gpc^2\,ms}}}{\nucounit^{1/3}P^{1/3}(r_c/\rlc)^{1/3}
\sqrt{C(\omega/\omco)f(\xi)}}
\nonumber\\
E&\gtrsim&\frac{1.63\times 10^{37}{\,\rm ergs\,}\epse\nucounit^{2/3}
(\Lnu/{\rm Jy\,Gpc^2\,ms})}
{P^{1/3}(r_c/\rlc)^{1/3}C(\omega/\omco)f(\xi)}~.
\label{QEburst}
\ea
Eqs. (\ref{QEburst}) do not depend on $N_i$: they follow from the total luminosity of the burst.

If $\mathscr{V}_i$ is the volume occupied by the incoherently superposed clumps, then the characteristic
charge and energy reservoirs associated with this region are given by Eqs. (\ref{Qcr}) and (\ref{EBr}) with the replacement
$C\mathscr{V}_s\to \Vi$:
\ba
Q_i&=&\frac{\Omega\mu\mathscr{V}_i}{2\pi ecr^3}\sim\frac{\mathscr{V}_i}{2\pi er^3}\sqrt{\frac{cI\dot P}{P}}
=10^{31}I_{45}^{1/2}(\dot P/P)_3^{1/2}(\mathscr{V}_i/r^3)
\nonumber\\
E_B(r)&=&\frac{\mu^2\mathscr{V}_i}{8\pi r^6}\sim\frac{I(\dot P/P)\mathscr{V}_i}{4P(r/\rlc)^3r^3}
=\frac{7.9\times 10^{33}\,{\rm ergs}\,(\dot P/P)_3(\mathscr{V}_i/r^3)}{P(r/\rlc)^3}~.
\label{reservoirs}
\ea
Superficially, the first of Eqs. (\ref{QEburst}) is well within the charge reservoir $Q_i$, thus allowing $\Vi/r^3$
to be small. However, recall that $Q_b^2$ is the {\sl total squared charge} for $N_i$ clumps, so if all of the $Q_c$ 
have the same sign then the total charge involved in the outburst is $\sim Q_b\sqrt{N_i}$, which would almost certainly
exceed the available charge $Q_i$ even if $\mathscr{V}_i/r^3\sim 1$. Curvature radiation for individual clumps does
not depend on the sign of their charges, though, and this difficulty is avoided if positive and negative $Q_c$ are
equally likely, in which case the expected net charge would be $\sim\pm Q_c\sqrt{N_i}\sim \pm Q_b$.

The energy required by the second of Eqs. (\ref{QEburst}) is harder to reconcile with Eq. (\ref{reservoirs}); taken
together they require
\be
\frac{\Lnu}{\rm Jy\,Gpc^2\,ms}\lesssim\frac{1.6I_{45}(\dot P/P)_3(r_c/\rlc)^{2/3}\nuunit^{1/3}C(\omega/\omco)f(\xi)}
{(P/30\,{\rm ms})^{1/3}[\epse(\omega r_c/c)^{1/3}]\nucounit^{2/3}}
\left(\frac{\Vi\rlc^3}{r^6}\right)~,
\label{Lnulim}
\ee
where we have adopted the most favorable case in which outbursts are associated with the first 
$\sim 1000$ years of a pulsar's life, when it might be expected to resemble the Crab pulsar.
Eq. (\ref{Lnulim}) favors (i) small $\epse\simeq(c/\omega r_c)^{1/3}$, its minimum value, 
(ii) large $\dot P/P$, (iii) short $P$ and (iv) comparatively small distances,
\be
\Dgpc\sqrt{\frac{\Snuth W_i}{\rm Jy\,ms}}\lesssim\frac{1.2 I_{45}^{1/2}(\dot P/P)_3^{1/2}(r_c/\rlc)^{1/3}\nuunit^{1/6}\sqrt{C(\omega/\omco)f(\xi)}}
{(P/30\,{\rm ms})^{1/6}[\epse(\omega r_c/c)^{1/3}]^{1/2}\nucounit^{1/3}}
\left(\frac{\Vi\rlc^3}{r^6}\right)^{1/2}
\label{dlim}
\ee
where $\Snuth$ is the detection threshold. 
If we estimate $\mathscr{V}_i\simeq(cW_i)^3$ then $\mathscr{V}_i/\rlc^3\simeq (\Omega W_i)^3
=9.2\times 10^{-3}[W_{i,{\rm ms}}/(P/30\ {\rm ms})]^3$. Using this as a benchmark value in
Eq. (\ref{dlim}) implies
\be
\Dgpc\sqrt{\frac{\Snuth W_i}{\rm Jy\,ms}}
\lesssim
\frac{0.11W_{i,{\rm ms}}^{3/2}I_{45}^{1/2}(\dot P/P)_3^{1/2}(r_c/\rlc)^{1/3}\nuunit^{1/6}
\sqrt{C(\omega/\omco)f(\xi)}}{(P/30\,{\rm ms})^{5/3}[\epse(\omega r_c/c)^{1/3}]^{1/2}\nucounit^{1/3}}
\left[\frac{\Vi\rlc^3}{r^6(\Omega W_i)^3}\right]^{1/2}
\label{dlimp}
\ee
Eq. (\ref{dlimp}) depends sensitively on $P$ and $r$, and less sensitively 
on $\dot P/P$. Pulsars younger and faster spinning than the Crab pulsar 
would be visible to larger distances: for 
$(\dot P/P)_3^{1/2}(P/30\,{\rm ms})^{-5/3}(\rlc/r)^3\simeq 10$ 
the limiting distance implied by Eq. (\ref{dlimp}) is $\simeq 1$ Gpc.
However, ``ordinary'' pulsars with $\dot P/P=10^{-7}(\dot P/P)_7\,{\rm y^{-1}}$
are visible to $\sim 3(\dot P/P)_7^{1/2} P^{-5/3}(\rlc/r)^3$ kpc, where $P$ is in
seconds. This is within the Galaxy for $\rlc/r\simeq 1$, and to be visible
at $\simeq 1$ Gpc emission would have to originate at $r/\rlc\lesssim 0.01$.
Single pulses with amplitude $\sim 0.2$~Jy and widths $\sim 20$~ms have been seen from pulsar J0529$-$6652 ($P= 975$~ms, $\dot P/P \sim5\times 10^{-5}$~yr$^{-1}$) in the Large Magellanic cloud
with $d = 54$~kpc \citep[][]{2013ApJ...762...97C}. 
Magnetars have $(\dot P/P)_3\sim 0.01-1$
and $P\sim 1-10$ seconds, and would therefore be visible to 
$\lesssim 300(\dot P/P)_3^{1/2}P^{-5/3}(\rlc/r)^3$ kpc, normalizing to the
{\sl most favorable} values of $\dot P/P$ and $P$; for $\rlc/r\simeq 1$
this limiting distance is well within the Local Group, and for typical
magnetar parameters, within the Galaxy. For detection to $\simeq 1$ Gpc 
radiation would have to originate at $r/\rlc\lesssim 0.01-0.1$.

Henceforth, we tentatively assume that the coherently emitting clumps that comprise an ERB resemble the clump that
caused the Crab's largest giant shot pulse, and use the properties we have inferred for it to consider other constraints.

\subsection{Radiation Drag}

The clump experiences drag by ambient radiation in addition to the ``radiation reaction drag'' it experiences
by interactions with its own radiation field. The drag force depends on how opaque the clump is, which requires us to
model its shape, but at a given $N$ an upper bound is found by assuming that the clump is optically thin, which
implies that the clump loses energy via drag at a rate $-\dot E_{drag}\lesssim N\gamma^2c\sigma_Tf(k)U_{rad}$, where 
$U_{rad}$ is the ambient radiation energy density in the inertial frame where the clump moves relativistically. The
factor $f(k)$ depends on the characteristic photon energy $km_ec^2$ in the rest frames of scattering leptons in the clump: 
$f(k)<1$ in general, $f(0)=1$ and $f(k)\simeq 3\ln(2k)/8k$ for $k\gg 1$ \citep[e.g.][]{1976tper.book.....J}.
Since the clump has energy $N\gamma m_ec^2$, the drag lifetime is $t_{drag}$, where
\be
\frac{ct_{drag}}{r_c}\gtrsim\frac{m_ec^2}{\gamma r_c\sigma_Tf(k)U_{rad}}
=\frac{5.61\times 10^4(\omco/\ompk)^{1/3}P^{11/3}}{\nucounit^{1/3}(r_c/\rlc)^{4/3}I_{45}\dot P_{15}(f(k)U_{rad}/U_{lc})}~;
\label{tdrag}
\ee
we estimated $U_{rad}$ by $U_{lc}=I\Omega\dot\Omega/4\pi\rlc^2c=4\pi^3I\dot P/c^3P^5
=4.59I_{45}\dot P_{15}P^{-5}\,{\rm erg\,cm^{-3}}$, where $I=10^{45}I_{45}\,{\rm g\,cm^2}$ is 
the moment of inertia of the neutron star, and $\dot P=10^{-15}\dot P_{15}$ is the spin-down rate. Although Eq. (\ref{tdrag})
suggests that drag is unimportant for nominal parameter values, this may be deceptive in specific cases: for Crab pular
parameters, Eq. (\ref{tdrag}) implies 
\be
\frac{ct_{drag}}{r_c}\gtrsim\frac{5\times 10^{-4}(\omco/\ompk)^{1/3}}{\nucounit^{1/3}(r_c/\rlc)^{4/3}I_{45}f(k)(U_{rad}/U_{lc})}
~~~~~~~~~~~~[{\rm Crab}]
\label{dragcrab}
\ee
suggesting a short lifetime: for comparison,  
$\epse\gtrsim 3\times 10^{-3}\nuunit^{-1/3}(r_c/\rlc)^{-1/3}$ for the Crab according to Eq. (\ref{epserange}), 
comparable to but larger than Eq. (\ref{dragcrab}).
The mild discrepancy may be resolved if the clump is moderately opaque even if $k\ll 1$; and the drag timescale is considerably longer
-- perhaps even $\sim r_c/c$ -- if $k\gg 1$.

Within the larger volume $\Vi$, coherent regions can interact with one another via their radiation
fields. If $W_i\sim 1$ ms and the radiation-reaction lifetimes of clumps are 
$r_c/c\gamma\simeq 0.15P/\gamma_3$ ms (i.e. near the small end of Eq. (\ref{epserange})
then the radiation energy density within $\Vi$ during the lifetime of a
particular coherently emitting clump is within an order of magnitude of the total energy emitted during the burst.
The radiation density is in the form of a concentrated beam of energy directed toward the observer. This directed
energy would accelerate the clumps. A detailed calculation of this effect requires more specific modelling of the
overall emission event, and we do not attempt it here.

\subsection{Pair Annihilation}

Since the clump consists of $N$ electrons and positrons with a relatively small net excess of one charge $Q_c\ll N$
its lifetime may also be limited by $e^\pm$ annihilation. If $n_{\rm rf}$ is the pair density in the rest frame of the
clump, then the annihilation rate per volume, a Lorentz scalar, is $\sim n_{\rm rf}^2(\sigma v)_{ann}$, where
$(\sigma v)_{ann}\approx\pi e^4A(\gint)/m_e^2c^3=\frac{3}{8}\sigma_TcA(\gint)$
is the pair annihilation rate coefficient for relative motion with Lorentz factors $\gint$,
where $A(1)=1$ and $A(\gint)\simeq \ln\gint/\gint^2$ for $\gint\gg 1$
\citep[e.g.][]{1976tper.book.....J}.
(We assume that the clump is hot enough in its rest frame that we can neglect annihilation via the formation of
positronium.) 
Then
Lorentz transform to the inertial frame of the magnetosphere, where the density is $n_b=\gamma n_{\rm rf}$, to get  the annihilation rate per lepton, 
$n_b(\sigma v)_{ann}/\gamma^2$,
implying a lifetime
$t_{ann}$, where
\be
\frac{ct_{ann}}{r_c}=\frac{\gamma^2c}{n_br_c(\sigma v)_{ann}}
\sim\frac{8\gamma^2(c/\omco r_c)}{3\tau_bA(\gint)}\gtrsim\frac{8(\omco r_c/c)^{2/3}(\ompk/\omco)^{2/3}}
{3\tau_b}
\ee
where $\tau_b$ is the Thomson optical depth of the clump and the length of the clump along
its direction of motion is $\sim c/\omco$. The lifetime of the clump against pair annihilation is long compared
with its drag and radiation reaction lifetimes.

\subsection{Conjectures on Clumping Mechanisms}

A possible mechanism for the formation of the relativistic clump responsible for the sub-bursts
is the bunching instability associated with coherent synchrotron radiation. There has not yet been
a treatment of this instability that made specific reference to a nearly neutral system with
$e^\pm$ pairs: analyses done to date \citep{1971ApJ...170..463G,2005PhRvE..71d6502S,2006PhRvS...9k4401S}
considered configurations of relativistic electrons only. Linear growth rates found in these treatments
appear to be fast enough to initiate the instability on timescales much shorter $\delta t_e$ on small
length scales.
However, we have seen that coherent emission requires a large number of leptons confined
to a small volume, leading to extremely large density contrast relative to the 
GJ density; thus a nonlinear analysis is needed to establish that this
mechanism is viable.
Another possible mechanism for generating dense clumps is the two stream instability
\citep[e.g.][]{rs75, 1977MNRAS.179..189B, 1978MNRAS.185..493B}.  
Magnetospheric reconnection events conceivably may play a role by producing relativistic  
charge streams or could trigger secondary activity (e.g. pair production) that leads to giant bursts.

\section{Extragalactic Populations of Neutron Stars}
\label{sec:pops}

Extragalactic NS  formed over cosmological time are potential sources of 
super-strong bursts that occur only rarely per object.   
We associate the burst rate with the cosmological birth rate of NS,  each object emitting
a small number of detectable bursts,   $\NGone = \eta_b T_b$,    at a low rate
 $\eta_b$ in a burst  phase of duration $T_b$.    A more detailed calculation could include separate
 birth  and burst rates, but the net result  is the same.  
 We use  the Galactic rate of NS formation per unit stellar mass, 
$\BRminline \approx \BRmFiducial10^{-13}~\rm yr^{-1}\, \Msun^{-1}\, $ 
(e.g. one NS every 100 yr per $10^{11}~\Msun$ in stars) in combination with the average stellar mass
density $\rho_\star = \Omega_\star \rho_c$ to calculate the aggregate NS birth and ERB burst rates, where $\rho_c = 3H_0^2/8\pi G$ is the closure density
and $\Omega_\star \approx 0.003 \OmstarFiducial$ \citep[][]{2005RSPTA.363.2693R}. 

About  $\NNS \sim 10^{17}$ NS are produced  in a Hubble volume $V_H = 4\pi d_H^3/3$ 
for a Hubble distance  $d_H = c/H_0 = 4.3h_{0.7}^{-1}  $~Gpc and a  typical
galaxy age $T_{\rm gal} = 10$~Gyr.
The aggregate NS birth rate scaled from the local rate in the Galaxy, 
\be
\Birthrate =   \rho_c \Omega_*  \BRm V_H
	 \sim 4\times 10^4~\rm day^{-1}\,  h_{0.7}^{-1}\, \BRmFiducial\, \OmstarFiducial,
\ee
is tantalizingly similar to the inferred ERB burst rate 
$\burstrateobs \approx 10^3$-$10^4~{\rm day^{-1}}$ \citep[][]{2013Sci...341...53T, 2014ApJ...790..101S, 2015ApJ...807...16L, 2015MNRAS.447.2852K}, suggesting that only of order one burst per NS is needed to account for ERBs. 
Larger $N_b$  would allow a nearer population with $\dmax \ll d_H$.

In the following we estimate ERB rates by including beaming of radiation and a redshift-dependent star formation rate (SFR).
  We assume that beaming toward Earth is favorable for a 
fraction  $f_b = 0.1f_{b,0.1}$ of the  bursts that are bright enough to be
detectable throughout the relevant volume.  Values of $f_b$ for pulsars are similar but depend  on spin period and on the emission being steady as the beam rotates across the sky.  ERBs could have substantially smaller $f_b$ if burst durations are a purely
temporal phenomenon in a frame rotating with the pulsar.  However,  the  beaming
fraction  for $f_b$ can be  comparable to that of pulsars if pulse widths are mostly due to beam rotation. 
The aggregate burst rate per unit volume is then $d\burstrate/d{\rm (vol)} = f_b \NGone \BRminline \rho_\star$.

\subsection{Implications of Non-detections in Followup Observations}
\label{sec:reobs1}

\newcommand{\rateperNS}{\eta_{\rm b}}

Individual ERB fields have been reobserved for up to $\sim 40$~hr, yielding no repeat bursts 
\citep[][]{2007Sci...318..777L}  but providing identification of one new ERB
\citep[][]{2015MNRAS.447..246P, 2015MNRAS.454..457P}.  One conclusion that can be made is
that the population is extragalactic. 
 This may be seen by calculating the burst rate {\it per NS} as a function of maximum population distance, $\rateperNS = \burstrateobs (d_H/\dmax)^3 / f_b \Birthrate T_{\rm gal}$. 
 To have $\rateperNS \lesssim 1$~d$^{-1}$ requires $\dmax \gtrsim 0.4$~Mpc.  
 However, if bursts are associated with a special phase in a NS's lifetime, the available subset of NS is much smaller, requiring more bursts per NS and accordingly a larger maximum distance to avoid any repeats.

\newcommand{\FoV}{\Omega_{\rm FoV}}
\newcommand{\FoVdegsq}{\Omega_{\rm FoV, deg^2}}

The lack of redetections also provides constraints on the burst rate per source $\eta_b$ depending on assumptions about the pulse-amplitude PDF.   For simplicity here, we assume that all bursts have the same intrinsic luminosity and can be detected out to some maximum distance $\dmax$ (also discussed as Case I in Section~\ref{sec:Nbconstant}).  The sky density of such sources is  $n_{\Omega}$~sources~deg$^{-2}$. 

Let  $\FoV$ be the solid angle of the telescope field of view (FoV)  for a single pointing and consider a total followup time  $T_F$ for an individual sky position.     Assuming many ERB sources are within the FoV, $f_b n_{\Omega}\FoV \gg 1$, where $f_b$ is  the beaming factor, as above,  the mean number of detected bursts is
\be
\langle N_{\rm ERB}\rangle =   \rateperNS T_F\,  (f_b\, n_{\Omega}\, \FoV).
\ee
If sources are sparsely distributed on the sky (or highly clustered) so that only the originally detected ERB source is within the FoV,   $f_b  n_{\Omega} \FoV$ would be replaced by unity, so 
generally $\langle N_{\rm ERB}\rangle =   \rateperNS T_F\,  \max(1, f_b\, n_{\Omega}\, \FoV)$.
With no redetections, $\langle N_{\rm ERB}\rangle \ll 1$ and  the limit on the rate is 
\be
\rateperNS  \ll   \frac{1}{T_F \max(1, f_b n_{\Omega} \FoV)}, 
\ee
which  shows   that the rate per source derived from reobservations  needs to take into account the sensitivity to {\it all}  ERB sources in the FoV that are sampled simultaneously, not just the source of the original ERB.   

The number of  extragalactic NS in the FoV of  any single-dish telescope is very large.   
With $N_{\rm NS} \sim 10^{17}$ NS in a Hubble volume, the sky density 
for $\dmax = 1~{\rm Gpc}~d_{\rm Gpc}$ is
\be
n_{\Omega} = \frac{N_{\rm NS}}{4\pi} \left(\frac{\dmax}{d_H} \right)^3
	\approx 3\times 10^{10}\, d_{\rm Gpc}^3 \, {\rm deg}^{-2} .
\ee 
To have only a single ERB source in a 1~deg$^{-2}$ FoV would imply a local population of
NS with a very small  population distance,
\be
\dmax \le 0.7~{\rm Mpc} \, h_{0.7}^{-1} \left( f_{b, 0.1}\,\FoVdegsq\right)^{-1/3}, 
\ee
which is inconsistent with the sky locations of ERBs seen to date. 
The implied rate per ERB is therefore very small for reobservations done so far,
\be
\rateperNS  \ll 3 \times 10^{-7}~{\rm hr}^{-1} \,  
\left(f_{b, 0.1}\,T_{\rm F, hr}\, \FoVdegsq\right)^{-1} d_{\rm Gpc}^{-3},
\ee
even taking into account $\FoV \sim 0.06$~deg$^2$ for a single beam of the Parkes telescope, 
which is nearly cancelled by $T_{\rm F} \sim$~tens of hours reobservation times.
The very small rate limit is completely consistent with the statement that only a small number of bursts is needed per NS if the population extends to Gpc distances. 

\subsection{Low-redshift Population}

For a local (low-redshift) population, the burst rate for a maximum population distance $\dmax$ is
$\burstrate(\dmax) = \burstrateH \left(\dmax/d_H\right)^3$, where the  local burst rate is normalized to a Hubble volume, 
\be
\burstrateH
	= 
	f_b \, \NGone\, \Birthrate
	=
	\frac{c^3 f_b\, {\NGone}_0\, \BRminlinezero\, \Omega_{\star,0}}
	       {2 H_0 G}
	\approx 4\times10^3~\rm day^{-1} ~  f_{b, 0.1}\,   \NGone  
		\left(\frac{\Birthrate}{ 4\times 10^4~{\rm day}^{-1}}\right)~.
\label{eq:gammaH}
\ee
To match  $\burstrate(\dmax)$ with 
 $\burstrateobs$ 
  requires   
\be
\NGone = \frac{\burstrateobs}{f_b\Birthrate } \left(\frac{d_H}{\dmax} \right)^3
	\approx  \frac{ 200}{f_{b, 0.1} d^3_{\rm max, Gpc}} 
	\left(\frac{\burstrate}{ 10^4~{\rm day}^{-1}}\right)
	\left(\frac{ 4\times 10^4~{\rm day}^{-1}}{\Birthrate}\right)
	~.
\ee
Even for a relatively nearby population with $\dmax = 100$~Mpc, the number of  bursts per object,
$\NGone \approx 2\times 10^5$, is a small fraction of the total number of turns of a NS  in
its lifetime, whether defined as the giant-pulse emitting time span, the total radio-emitting span,
or the age of the NS itself.  
In time $T_b$ the number of turns  is $\phi_T \sim  \dot P_0^{-1}( T_b / \tau_0)^{(n-2)/(n-1)}$ for a
starting period $P_0$ and period derivative $\dot P_0$, yielding a  spindown time $\tau_0 = P_0 / (n-1)\dot P_0$ for a braking index $n$.   
 The Crab pulsar will make $\sim 10^{12}$  turns  in the next $10^3$~yr  and  $10^{14}$ turns in $10^9$~yr
 assuming a constant braking index.   If ERBs are similar to giant pulses emitted in the first $T_b = 10^3$~yr of a NS,  the interval between ERBs per object is only  tens to hundreds of hours.    However, rare events occurring at any point in the lifetime of a NS (e.g. $\gtrsim 1$~Gyr) imply intervals of tens to hundreds of years or longer.  
 
The number of bursts per source  is tied to the contributions of the IGM and any host galaxy to  the observed DMs of ERBs.  If the maximum population distance is
of order $d_H$, the IGM can and must account for nearly all the excess over the Milky Way's contribution.  Conversely,  a nearby $\sim 100$~ Mpc (or closer)  population requires a dominant contribution from
host galaxies,  in particular from dense star-forming regions or galactic centres.

\subsection{High-redshift Population}

For  populations extending to high redshifts, we relate the local NS birth rate $\BRminline(z)$ to  the SFR $\psi(z)$
\citep[in standard units $\Msun~\rm yr^{-1}~Mpc^{-3}$; e.g.][]{2014ARA&A..52..415M}
\be
\ratepermass\rhostar=\nuns\psi(z)
\ee
where $\nuns$ is the number of neutron stars per stellar mass formed at redshift $z$
and the stellar mass density is a co-moving density.  
Using the beaming fraction, assumed independent of $z$,  and the number of bursts per NS,
$\NGone(z)$, now possibly redshift dependent,  the burst rate out to a maximum redshift $\zmax$ is
\ba
\Gamma_b(\leq \zmax)&=&4\pi f_b\int_0^{\zmax} dz\,\frac{dr(z)}{dz}\frac{r^2(z)N_b(z)\nuns\psi(z)} {(1+z)}~.
\ea
where $r(z)$ is the comoving radial coordinate and $4\pi dz\,r^2(z)dr(z)/dz$ is the comoving volume element. 
The $1+z$ factor in the denominator accounts for the redshift of the NS birth rate. 

By normalizing $\nuns$ and $\psi(z)$ to local values,
\be
\nuns =  \nunszero\nuhns \quad {\rm and} \quad \psi(z) = \psi(0) \psihat,
\ee
the rate becomes
\be
\Gamma_b(\leq \zmax) = 
4\pi f_b \BRminlinezero\rhostarnow\int_0^{\zmax} dz\,\frac{dr_c(z)}{dz}\frac{[r_c(z)]^2N_b(z)\nuhns\psihat}{1+z}~.
\label{eq:burstrate}
\ee

\subsubsection{Case I: Constant $N_b$ vs. $z$}
\label{sec:Nbconstant}

One possibility is that ERBs are associated with extremely rare,  triggered events occurring $N_b$ times during the lifetime of a NS.  
If bright enough intrinsically -- e.g.  flux densities $\gtrsim 10^{12}\,{\rm Jy}$ at a fiducial 
kpc distance -- such
events could be detectable at cosmological distances, up to some cutoff $z_c\gtrsim 1$. 
In this case  $N_b(z)$  is independent of $z$ since it is determined entirely by activity internal to the NS or in the magnetosphere, or to external triggers that do not involve redshift-dependent
conditions. 
  In general,  $\nuns$ depends on the initial mass function (IMF), which depends
on $z$. However, at modest $z\lesssim 1$ (but not $z\gg 1$) it is also plausible
that it is close to the local value and therefore $\nuhns = 1$.   
Under these circumstances we write the burst rate in terms of a dimensionless integral $G_b(z)$  over the normalized SFR, $\hat\psi(z)$, 
\be
\Gamma_b(\leq z_{\rm max}) &=& \burstrateH G_b(\zmax), 
\quad\quad \text{with} \quad
G_b(z) = 3\int_0^{z} d\zp\,\frac{d\rtil(\zp)}{d\zp}\frac{\rtil^2(\zp)\psihatp}{1+\zp}~.
\ee
The dimensionless radial coordinate is
$\rtil(z) = \int_0^z dz^{\prime} / E(z^{\prime})$ where 
$E(z) = \sqrt{1 - \Omega_M + \Omega_M(1+z)^3}$ for a $\Lambda$CDM cosmology with
a fractional matter density $\Omega_M$. 

\begin{figure}
\begin{center}
\includegraphics[scale=0.39]{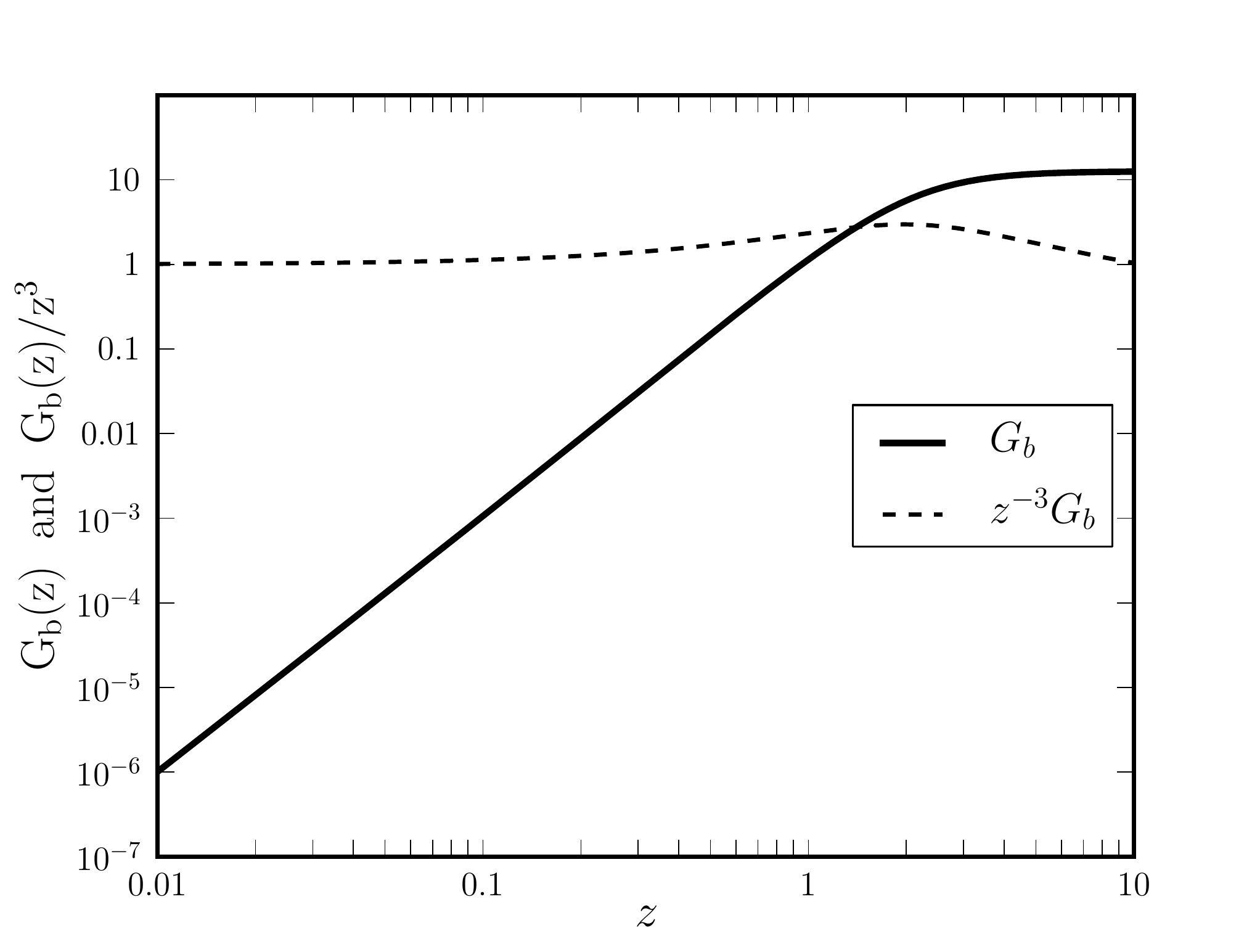}
\caption{Volume integral $G_b(z)$ for bursts in a  $\Lambda$CDM cosmology with $\Omega_M=0.27$
for the case where the number of bursts per source $\NGone$ is independent of $z$;   
}
\label{fig:madaurateplot}
\end{center}
\end{figure}

To obtain $G_b(z)$ we use   the empirical  fit to the cosmological SFR 
\citep[][Eq. 15]{2014ARA&A..52..415M}, 
\be
\psi(z) = 
\frac{0.015(1+z)^{2.7}}{1 + [(1+z)/2.9]^{5.6}}~\Msun~{\rm yr^{-1}~Mpc^{-3}},
\ee
and normalize it to $z=0$ to get $\psihat$. 

Figure~\ref{fig:madaurateplot}  shows $G_b(z)$.    
For low maximum redshifts,  $G_b(z) \approx z^3$ so
the ratio $G_b(z) / z^3$, also shown in the figure, is unity as expected.  
At high maximum redshifts $G_b(z)$ flattens to a value $\sim 10$ times the value at $z=1$,
indicating that a very distant population would not require all NS to contribute 
to observed ERBs\footnote{Dispersion measures of ERBs indicate redshifts $z < 1$ if the IGM is smoothly distributed.  The clumpiness of the cosmic web may allow more distant objects to be seen with the measured
DMs \citep[][]{2007Sci...318..777L, 2013Sci...341...53T, 2014ApJ...790..101S}.   However, scattering in the IGM or radiation physics may not allow bursts from such sources to be detectable.}.  


\subsubsection{Case II: $N_b(z)$ for an Intrinsic Distribution of Burst Luminosities}
\label{sec:highz}

\def\Lnu{{\mathscr{L}_\nu}}
\def\Lnumin{{\mathscr{L}_{\nu,{\rm min}}}}
\def\Lnuu{{\mathscr{L}_{\nu,u}}}
\def\Lnul{{\mathscr{L}_{\nu,l}}}
\def\Lnuone{{\mathscr{L}_{\nu,1}}}
\def\Anuth{A_\nu^{({\rm th})}}
\def\Lnuth{\Lnu^{({\rm th})}}
\def\LnuthH{\mathscr{L}_{\nu,H}^{({\rm th})}}
\def\Snuth{S_{\nu,{\rm th}}}
\def\dmin{d_{\rm min}}
\def\zmin{z_{\rm min}}
\def\Nbtotal{N_{b,{\rm total}}}
\def\Nscrb{{\mathscr{N}}_b}

In actuality, $N_b(z)$ is determined by the number of bursts per pulsar bright enough to
be detected to redshift $z$ via observations at frequency $\nu$. This involves the
distribution of intrinsic burst luminosities, $\Lnu$ (with units of luminosity per unit frequency). We express  the number of bursts with luminosities above $\Lnu$ as $\Nscrb(>\Lnu)$.

The key is detectability: we suppose that a burst will be detected with a flux density
threshold $\Snuth$ over a timescale $W_i$; we define $\Anuth=\Snuth W_i$. For a source at
redshift $z$, we refer the threshold luminosity to the observational frequency, so the 
$\Lnuth$ involves both an appropriate redshift-dependent distance and a color correction;
we find
\be
\Lnuth=\frac{P(\nu)d_L^2(z)\Anuth}{(1+z)P(\nu(1+z))}=\frac{P(\nu)(1+z)r^2(z)\Anuth}{P(\nu(1+z))}
=\frac{P(\nu)(1+z)\rtil^2(z)\LnuthH}{P(\nu(1+z))}
\ee
for a source at redshift $z$, where $P(\nu)$ is the spectrum; $d_L(z)=(1+z)r_c(z)$ is the luminosity distance
where $r_c(z)=d_H\rtil(z)$ is the radial coordinate to the source, and
\be
\LnuthH=\Anuth d_H^2=\frac{18.4\,{\rm Jy\,ms\,Gpc^2}}{h_{0.7}^2}\left(\frac{\Anuth}{1\,{\rm Jy\,ms}}\right)
=\frac{18.4\times 10^{12}\,{\rm MJy\,ns\,kpc^2}}{h_{0.7}^2}\left(\frac{\Anuth}{1\,{\rm Jy\,ms}}\right)
\label{LnuthHdef}
\ee
where $d_H=c/H_0=4.29h_{0.7}^{-1}$ Gpc for $H_0=70h_{0.7}\,{\rm km\,s^{-1}\,Mpc^{-1}}$.
For coherent curvature radiation, $P(\nu)\propto\nu^{2/3}$ below a cutoff frequency $\nu_c$ above which the
spectrum cuts off exponentially so we approximate
\be
f(z)=\frac{P(\nu)(1+z)\rtil^2(z)}{P(\nu(1+z))}\simeq \rtil^2(z)(1+z)^{1/3}e^{\nu z/\nu_c}\equiv \rtil^2(z)
(1+z)^{1/3}e^{z/z_c}
\label{color}
\ee
where we bear in mind that $z_c\equiv\nu_c/\nu$ depends on the observational frequency $\nu$. Notice
that $f(z)$ is an increasing function of $z$: as $z$ increases, the sources that we can detect have
ever increasing intrinsic luminosities. With these definitions, 
\be
N_b(z)=\Nscrb(>\LnuthH f(z))~,
\label{Nbofz}
\ee
which can be inserted into Eq. (\ref{eq:burstrate}).

Eq. (\ref{eq:burstrate}) integrates the rate to $\zmax$. Via energetic arguments, we have shown that 
there is an upper luminosity cutoff $\Lnuu$ to the distribution of bursts. Consequently, for a given 
observational survey, $\zmax$ is determined by solving $\Lnuu=\LnuthH f(\zmax)$. There is also a minimum
burst luminosity, $\Lnul$, although its definition may be problematic because it involves qualitative
criteria for distinguishing an outburst from ``normal fluctuations'' in pulsar luminosity. Nevertheless,
there is also a maximum redshift $z_l$ to which we can observe the faintest outburst, which is determined by
$\Lnul=\Lnuth f(z_l)$; all outbursts are detectable out to $z_l$, but only sufficiently luminous ones are
detectable for $z_l<z\leq\zmax$. Thus, the observed burst rate is
\be
\Gamma_b(\leq \zmax(\Snuth))=f_b\Gamma_{\rm ns}\left[\Nbtotal G_b(z_l(\Snuth))+\int_{z_l(\Snuth)}^{\zmax(\Snuth)}
dz\frac{dG_b(z)}{dz}\Nscrb(>\LnuthH f(z))\right]~,
\label{rateintegralth}
\ee
where $\Nbtotal$ is the {\sl mean}
total number of outbursts per pulsar brighter than $\Lnul$. 
Presumably
there is a distribution of $\Nbtotal$ that depends on pulsar properties, but we assume that this distribution
does not vary with $z$. 
The observed bursts will be dominated by the intrinsically {\sl brightest}
bursts if $\Nscrb(>\Lnu)$ is {\sl shallow}, and by the intrinsically {\sl faintest} bursts if it is {\sl steep}.
Given the uncertainty in $\Nscrb(>\Lnu)$, we define an effective number of bursts per pulsar as a function of $\zmax$ via 
$\Gamma_b(\leq \zmax)\equiv f_b\Gamma_{\rm ns}N_{b,{\rm eff}}(\zmax)G_b(\zmax)$, or
\be
N_{b,{\rm eff}}(\zmax(\Snuth))=\frac{1}{G_b(\zmax(\Snuth))}\left[\Nbtotal G_b(z_l(\Snuth))+\int_{z_l(\Snuth)}^{\zmax(\Snuth)}
dz\frac{dG_b(z)}{dz}\Nscrb(>\LnuthH f(z))\right]~.
\ee

\def\Nhscrb{\mathscr{{\hat N}}}
\def\Pscrh{{\mathscr{\hat P}}}

Let $\Nscrb(>\Lnu)=\Nbtotal\Nhscrb(>\Lnu)$ so that $\Nhscrb(>\Lnu_l)=1$ and $\Nhscrb(>\Lnu_u)=0$;
for a powerlaw distribution
\be
\Nhscrb(>\Lnu)=\frac{\Lnuu^{1-\alpha}-\Lnu^{1-\alpha}}{\Lnuu^{1-\alpha}-\Lnu_l^{1-\alpha}}~.
\ee
Measure the flux density relative to some reference value $S_0$; at this value, the threshold value of $\Lnu$
is $\LnuthH(S_0)$,  the faintest bursts are visible to some reference redshift $z_l(S_0)$, and the brightest 
bursts are visible to some higher reference redshift $\zmax(S_0)$. Then at any other
flux density threshold $S$, 
\be
f(z_l(S))
=\frac{f(z_l(S_0))}{S/S_0}
~~~~~f(\zmax(S))=\frac{f(\zmax(S_0))}{S/S_0}~;
\label{zrelations}
\ee
Eqs. (\ref{zrelations}) can be inverted to find $z_l(S)$ and $\zmax(S)$ {\sl given} $z_l(S_0)$ and $\zmax(S_0)$, 
and these relations could be differentiated with respect to $S$.

Since we are interested in how the burst rate varies with flux density threshold $S$, we denote $\Gamma_b(\leq\zmax(S))$
simply by $\Gamma_b(\geq S)$.
In Euclidean space with a $z$ independent star formation rate, 
$dG_b(z)/dz=z^2$ and $f(z)=z^2$, and $\Gamma_b(\geq S)\propto S^{-3/2}$ independent
of the distribution of intrinsic brightnesses, as is well known.  depends on a number of parameters, even after fixing 
cosmological parameters and the star formation rate: $z_l(S_0)$, $\zmax(S_0)$, $z_c$,
and $\alpha$. There are basic trends associated with these parameters:
\begin{itemize}
\item High $\alpha$ with low $z_l(S)$ implies  $\Gamma_b(\geq S)$ 
is dominated by intrinsically faint bursts which, in turn, implies that $S^{3/2}\Gamma_b(\geq S)$ remains nearly constant
for a wide range of $S$ even when $\zmax(S)$ is large.
\item For low $\alpha$,  $\Gamma_b(\geq S)$ is dominated by bursts that are both intrinsically bright  and have the largest flux densities, so  $S^{3/2}\Gamma_b(\geq S)$ drops
substantially at low $S$ for $\zmax(S)\gtrsim 1$.
\item Deviations from constant $S^{3/2}\Gamma_b(\geq S)$ arise as a consequence of curved spacetime geometry,
variable star formation rate, and color corrections. The salient effects include:
\begin{enumerate}
\item $\rtil(z)$ asymptotes at large $z$, for example, $\rtil\to 2.28$ in
$\Lambda$CDM with $\Omega_M=0.27$;
\item a peak star formation rate at $z\approx 1.86$;
\item slow (approximately logarithmic) increase of $\zmax(S)$ at small $S$ for $z_c\approx 1$ as color corrections become
significant.
\end{enumerate}
At $z<1.86$, the rising star formation can cause $S^{3/2}\Gamma_b(\geq S)$ to grow.
\end{itemize}

\begin{figure}
\begin{center}
\includegraphics[scale=0.60]{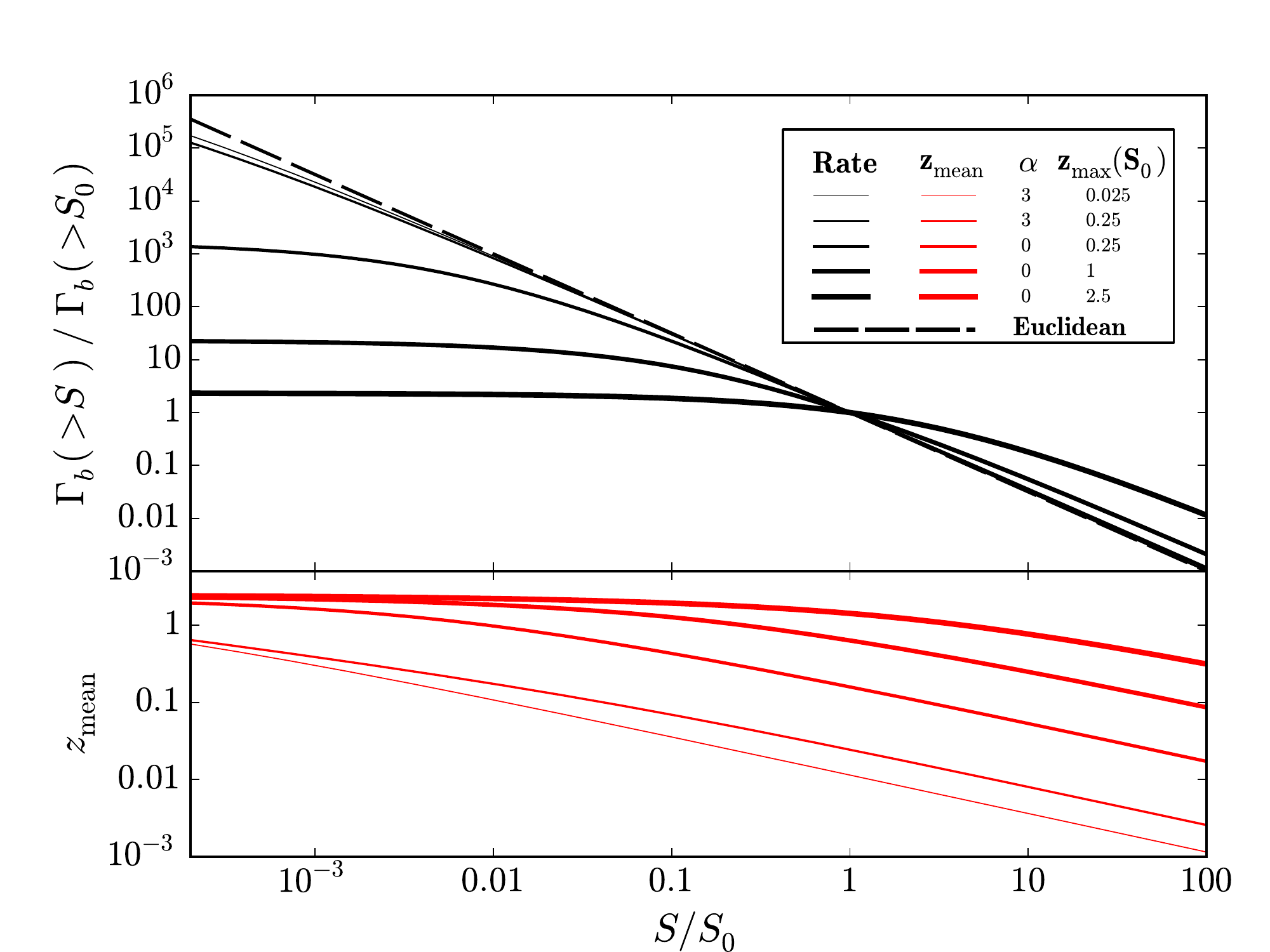}
\caption{
Plot of the scaled burst  rate and mean redshift $z_{\rm mean}$ for five combinations
of $z_{\rm max}$ and slope $\alpha$ of a power-law luminosity function for individual sources.
The rate is normalized to a value of unity at $S/S_0 = 1$.
The assumed spectrum has an exponential cutoff $z_c = 1$.  The dashed line shows
the $(S/S_0)^{-3/2}$ relation for a population of standard candles in Euclidean space.  
For more details see text (\S\ref{sec:highz})
\label{fig:fluxcdfs}
}
\end{center}
\end{figure}

The upper panel in Fig.~\ref{fig:fluxcdfs} shows a sample of results for 
$\Gamma_b(\geq S)/\Gamma_b(\geq S_0)$ assuming power law distributions
of intrinsic burst brightnesses. We regard $\Gamma_b(\geq S_0)$ as known observationally 
(e.g. $\sim 10^{3-4}$ per day at $S_0\simeq 1$ Jy). 
To aid in interpreting the results, 
the lower panel shows the mean redshift of outbursts 
\be
z_{\rm mean}(S)\equiv\frac{1}{\Gamma_b(S)}\int_0^{\zmax(S)} dz\,z\,\frac{d\Gamma_b(z,S)}{dz}~.
\ee
We fix $z_l(S_0)=0.01$,
which means that the intrinsically faintest ERBs would only be detectable to a distance $\simeq 40$ Mpc
at the reference flux density $S_0$; we deliberately chose $z_l(S_0)$ to be large enough that bursts
from the Local Group only would occur rarely. We adopt $z_c=1$ 
since the most favorable condition for detection is for frequencies near but below the curvature radiation peak.
The dashed curve is the Euclidean result $(S_0/S)^{3/2}$, which is shown for comparison.
Results are shown for $\alpha=3$, a {\sl steep} distribution for which the observed bursts are primarily the
{\sl intrinsically faintest}, and $\alpha=0$, a {\sl flat} distribution for which the observed bursts are
primarily the {\sl intrinsically brightest}. For $\alpha=3$ we show results for $\zmax(S_0)=0.025$, for which
the intrinsically brightest bursts are only visible to a distance $\simeq$ 100 Mpc at the reference flux
density threshold $S_0$, and $\zmax(S_0)=0.25$,
for which the intrinsically brightest bursts are visible to a distance $\simeq$ 1 Gpc at $S_0$. As can be seen
from the figure, deviations from Euclidean are minor even at $S/S_0\lesssim 10^{-3}$. The lower panel
shows that $z_{\rm mean}(S)\lesssim$ a few tenths at all $S/S_0$ shown, even though $\zmax(S)$ (which is
not shown) is quite large ($\simeq 10$ for the lowest $S/S_0$). The results are dramatically different
for $\alpha=0$, where observations are dominated by the {\sl intrinsically brightest} bursts. To highlight
cosmological effects, we show results for $\zmax(S_0)=0.25,\,1$ and $2.5$. In all of these cases,
$\Gamma_b(S)$ levels off at small $S/S_0$; for $\zmax(S_0)=2.5$ the distribution is even substantially
flatter than Euclidean at large flux densities, $S_0<S<100S_0$. The lower panel shows that the rate
cuts off when $z_{\rm mean}(S)\gtrsim 1$. However, although $\zmax(S)\simeq 10$ for the lowest values
of $S/S_0$ shown, $z_{\rm mean}(S)$ does not grow rapidly at small $S/S_0$, primarily because of the
effects of the color correction and diminishing star formation rate.

Leveling off of $\Gamma_b(S)$ at small values of $S$ is a
distinctive signature of a cosmological population of ERBs. However, if, in particular,
$\alpha$ is fairly large, discerning deviations from the Euclidean distribution would require
large and deep samples of bursts.
This is because, for a steeply decreasing distribution of
intrinsic brightnesses, observations at a given flux density threshold are dominated by sources at
relatively low redshifts, where deviations from Euclidean geometry are smallest. Moreover, if the
ERBs {\sl already detected} are at typical redshifts $\simeq 1$, the rate of detection at smaller
flux density thresholds will only grow by a modest factor: for $\alpha=0$ and $\zmax(S_0)=2.5$. 
Fig.~\ref{fig:fluxcdfs} shows
that $\Gamma_b(S)/\Gamma_b(S_0)\lesssim 2$ for $S/S_0\geq 10^{-3}$.
Conversely, if the ERBs already detected are primarily ``local,'' at distances $\lesssim 100\,{\rm Mpc}-
1\,{\rm Gpc}$, then
$\Gamma_b(S)$ continues to rise rapidly with decreasing $S/S_0$ until cosmological effects intervene.

\subsubsection{Role of Gravitational Lensing}
\label{sec:lensing}

The above estimates  ignore gravitational lensing, which has been considered
by \citet[][]{2014ApJ...797...71Z}.  
Here we consider gravitational lensing by stars; the characteristic
Einstein radius for source and lens at $z\sim 1$ is  $R_E=\sqrt{4GM/cH_0}\approx 8.9\times 10^{16}{\rm cm}(M/\msun)^{1/2}$,
much larger than stellar radii. Amplifications are $A(b)\sim R_E/b$ for rays passing at impact parameter
$b\ll R_E$, and the maximum amplification is $A(R)\sim 1.3\times 10^6(M/\msun)^{1/2}(\rsun/R)$.
The duration of a lensing event with amplification $A$ is
$R_E/Av\sim 94A^{-1}(M/\msun)^{1/2}(v/300\kms)^{-1}$ years, where $v$ is a characteristic relative
speed between lens and source projected onto the sky.
Because we are interested in short-lived pulses, it is the
lensing probability that is relevant, not the lensing rate, which is appropriate for steady light sources, such as
the stars monitored in microlensing surveys.
The probability density for $A$ is $p(A)dA\sim\Omega_\star
A^{-3}dA$ at large $A$, 
\be
P(A)=Ap(A)\sim\frac{\Omega_\star}{A^2}
\approx\frac{0.003 \Omega_{\star, 0.003}}{A^2}
\ee
is a measure of the probability of lensing amplification $\sim A$ out to $z\sim 1$. 
Thus, the rate of bursts that
are visible because of amplification by gravitational lensing is
\be
P(A)\Gamma_{\rm b,H}
\sim
\frac{
	0.003\Omega_{\star,0.003}\Gamma_{\rm b,H}
	}
	{A^2}
=
\frac{
	0.003\Omega_{\star,0.003} \Gamma_{\rm b,obs}}
	{A^2}
\left(\frac{d_H}{d_{\rm max}}\right)^3~,
\label{lensedrate}
\ee
assuming that {\sl unlensed} bursts detectable only out to
distance $d_{\rm max}$ can account for the observed rate by
themselves. The amplification factor needed to render a burst
at $z\sim 1$ detectable is $\simeq (d_H/d_{\rm max})^2$,
assuming that the intrinsic burst brightness is the same as
those detected without lensing, so the 
fraction of bursts that are detectable from $z\sim 1$ as a
result of lensing is
\be
f_{\rm lens}\sim\frac{0.003\Omega_{\star,0.003}}{(d_H/d_{\rm
max})}
\ee
which is $\sim 10^{-4}$ for $d_{\rm max}=100$ Mpc. Nevertheless,
lensed ERBs from $z\sim 1$ may be observable of order once per
day.

\subsection{DM Contributions from Galaxy Clusters}
\label{clusters}

\def\Lnu{{\mathscr{L}_\nu}}
\def\Lnumin{{\mathscr{L}_{\nu,{\rm min}}}}
\def\Lnuu{{\mathscr{L}_{\nu,u}}}
\def\Lnuone{{\mathscr{L}_{\nu,1}}}
\def\Anuth{A_\nu^{({\rm th})}}
\def\Lnuth{\Lnu^{({\rm th})}}
\def\LnuthH{\mathscr{L}_{\nu,H}^{({\rm th})}}
\def\Snuth{S_{\nu,{\rm th}}}
\def\dmin{d_{\rm min}}
\def\zmin{z_{\rm min}}
\def\Nbtotal{N_{b,{\rm total}}}

As stated earlier, ERB host galaxies  can provide a significant fraction of the burst's DM and therefore play a strong role in establishing the ERB distance scale.   Here we discuss the contribution to DMs from galaxy clusters, either as sites for host galaxies or as intervening structures. 

\newcommand{\fbary}{f_{\rm bary}}

Following \cite{2010A&A...517A..92A}, we parameterize the electron column density in a cluster via the mass
$M_\eta$ of a region within which the mean density is $\eta$ times the closure density at the redshift of the cluster
(see also \cite{2007ApJ...668....1N} and \cite{2005RvMP...77..207V}). In this prescription, $R_\eta=[2GM_\eta/\eta
H^2(z)]^{1/3}$ and we estimate the electron column density to be
\be
DM_{\eta} \approx
\frac{3\fbary\eta H^2(z)R_\eta(1+X)}{16\pi G m_p}\approx 340h_{0.7}^{4/3}\,{\rm pc\,cm^{-3}}
\left(\frac{M_\eta}{10^{14}\msun}\right)^{1/3}
\left(\frac{\eta}{500}\right)^{2/3}\left(\frac{\fbary}{0.2}\right)\left(\frac{1+X}{1.75}\right)[E(z)]^{4/3}.
\label{dmeta}
\ee
where $\fbary$  is the baryon density fraction and $X$ is the H fraction by mass 
in the cluster gas; $\eta=500$ is commonly used in fitting data on SZ clusters with generalized NFW pressure profiles
(e.g. \cite{2015arXiv150201597P}). The characteristic dispersion measure implied by Eq. (\ref{dmeta}) is substantial,
and given that the electron density varies significantly within a cluster, DM values several times larger or
smaller than Eq. (\ref{dmeta}) may be encountered for different lines of sight.  Since
the escape velocity is 
\be
\sqrt{\frac{GM_\eta}{R_\eta}}\approx 780h_{0.7}^{2/3}\,{\rm km\,s^{-1}}\left(\frac{M_\eta}{10^{14}\msun}\right)^{1/3} 
\left(\frac{\eta}{500}\right)^{1/6}[E(z)]^{1/3},
\ee
neutron stars are unlikely to escape the galaxy cluster even if they are able to escape the galaxies where they were
born; moreover, 
$1\,{\rm Mpc}/1000\,{\rm km\,s^{-1}}=0.98$~Gyr, so even a fast-moving neutron star is
unlikely to cross a cluster during its lifetime as a pulsar, much less during its first $\sim 1000$ active years.
Using the $M_{500}-Y_{500}$ relationship from Eq. (5) in \cite{2014A&A...571A..29P} (see also Eqs. (25)-(27) in
\cite{2010A&A...517A..92A})
we find a DM$_{500}$-$Y_{500}$ relationship:
\be
{\rm DM}_{500}
\approx
740h_{0.7}^{0.50}\,{\rm pc\,cm^{-3}}
\left(\frac{Y_{500}} {10^{-4}\,{\rm arcmin^2}}\right)^{0.19}
\frac{[E(z)]^{1.2}}{1+z}
\left[\frac{\rtil(z)}{1+z}\right]^{0.37}~.
\label{DMY}
\ee
(The additional factor of $1+z$ arises from the redshift of the radiation frequency between the cluster and
observer.)
Using the approximate mass ($M_{500}$)-richness ($N_{200}$) relationship from \cite{2014MNRAS.438...78R},
we find a DM-richness relationship (see also \cite{2012ApJ...749...56B})
\be
{{\rm DM}_{500}}
\approx 
440\,{\rm pc\,cm^{-3}}
(N_{200}/40)^{0.35\pm0.04}~.
\label{DMN}
\ee
 The relationships underlying Eq. (\ref{DMN}) were derived for clusters
at modest redshifts ($z\lesssim 0.3$; \cite{2009ApJ...699..768R}), so we have not included detailed $z$ dependences
in Eq. (\ref{DMN}).

 \citet[][]{2014ApJ...780L..33M} has demonstrated that a large ERB sample can be used to probe the distribution of ``missing'' baryons in the universe by analyzing DM values as a function of proximity of lines of sight to galaxy clusters.  In the next section we consider the likelihood of such proximities and the role of clusters in establishing a distance scale for ERBs.

\section{Summary of Constraints on ERB Populations}
\label{sec:summary}

Here we consolidate  constraints on ERB source distances made in previous sections on the basis of  NS population numbers  and radiation physics in NS magnetospheres.  We also incorporate the likely contributions to ERB dispersion measures from galaxy clusters for  at least some of the ERBs. 
 
\begin{figure}
\begin{center}
\includegraphics[scale=0.6]{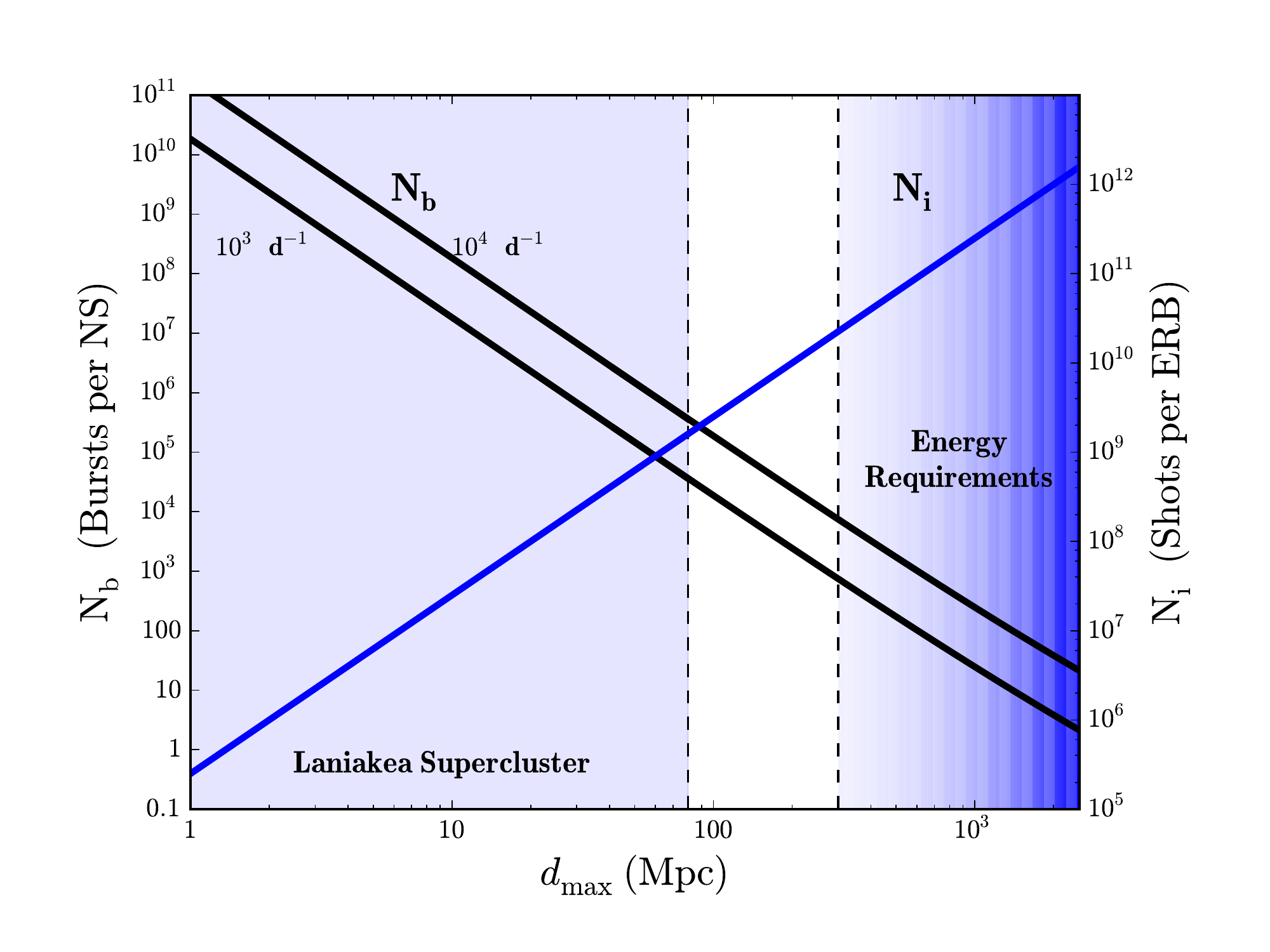}
\caption{ Plot showing the number of bursts per NS ($N_b$) and the number of shot pulses per ERB ($N_i$) needed to account for the apparent total sky rate of ERBs and their flux densities. Two curves are shown for $N_b$ based on ERB rates of $10^3$ and $10^4$~d$^{-1}$.  The shaded region 
on the left indicates the  approximate size of the Laniakea supercluster \citep[][]{2014Natur.513...71T}, which is approximately the  distance out to which there is about one L* galaxy per deg$^2$.  On the right, the increasingly darker shading indicates that   ERBs  challenge the energy requirements of coherent radiation.
}
\label{fig:phasespace}
\end{center}
\end{figure}

Figure~\ref{fig:phasespace} shows constraints on $N_b$, the number of bursts per NS (black lines), and on $N_i$, the number of shot pulses per burst needed to account for measured flux densities (blue line), as a function of maximum population distance.    The two lines for $N_b$ correspond to full-sky burst rates of $10^3$ and $10^4$~d$^{-1}$ and indicate that only a small number is needed per NS if the detectable population extends to $\sim 1$~Gpc.   For smaller $\dmax$, repeats may be expected if the burst phase of a given NS is short, such as the current phase of the Crab pulsar (1000~yr) but are not expected if ERBs occur predominantly from older, possibly dormant NS, that are $>$~Gyr in age. 
The number of shots per ERB increases rapidly with $\dmax$ and is based on assuming that individual shots are limited by radiation reaction and particle number densities expected in NS magnetospheres.  The shaded region on the right indicates that energy requirements become increasingly severe as  $\dmax$ increases from a ~few hundred Mpc to larger distances.   The shaded region on the left  represents  the approximate 80~Mpc radius of the local Laniakea supercluster  \citep[][]{2014Natur.513...71T},  comprising at least $10^5$ galaxies distributed in prominent galaxy clusters, including the Virgo cluster.  The corresponding galaxy density $\sim 2$~deg$^{-2}$, so the lack of associations of detected ERBs with any galaxies in the local supercluster suggests 
$\dmax \gtrsim 80$~Mpc.  

The unshaded region  in the figure between 80 and 300~Mpc  constitutes a plausible but by no means required range for $\dmax$.  
The distances in this range are near enough that cosmological effects are minimal, but far enough
  that it is unlikely that repeat bursts would be detected from any particular NS.

Additional constraints on the ERB distance scale involve the possible contributions of galaxy clusters to DMs. As shown previously, cluster DMs are  large enough to account for much of the DM in excess of the  contribution from the Milky Way's ionized gas.    However, we need to consider the likelihood for  an ERB source either residing  within or being viewed through a cluster.   
 For an overdensity $\eta$, a  cluster's angular size corresponding to $ R_\eta$  at redshift $z$ is
\ba
\theta_\eta&=&\left(\frac{2GM_\eta H_0}{\eta c^3}\right)^{1/3}\frac{1+z}{[E(z]^{2/3}\rtil(z)}=
\frac{0.57\,{\rm arcmin}\,
(1+z)}{(\eta/500)^{1/3}[E(z)]^{2/3}\rtil(z)}\left(\frac{M_\eta h_{0.7}}{10^{14}\msun}\right)^{1/3}
\nonumber\\
&\approx&\frac{2.4\,{\rm arcmin}
}{(\eta/500)^{1/3}h_{0.7}^{2/3}\Dgpc}\left(\frac{M_\eta}{10^{14}h_{0.7}^2\msun}\right)^{1/3}
\label{thetaeta}
\ea
where we used the low redshift approximation $\rtil(z)[E(z)]^{2/3}/(1+z)\simeq z
\simeq H_0d/c\approx 0.23h_{0.7}\Dgpc$ in the last line of Eq. (\ref{thetaeta}).
The expected number of cluster intersections made by a line of sight is 
\be
N_{\rm C, int}=\pi n_{C,0}\int dz\theta_\eta^2(z)\frac{dr_c(z)}{dz}r_c^2(z)
\ee
where $n_{C,0}=3f_C\Omega_mH_0^2/8\pi GM_C$ is the comoving density of clusters assuming that a fraction $f_C$ of the
mass of the Universe is in clusters of mass $M_C$; below, we take $M_C=M_{\eta}$ and $\eta=500$ for numerical estimates.
Using Eq. (\ref{thetaeta}) we find
\ba
N_{\rm C, int}&=&\frac{3f_C\Omega_mc}{4(2GM_\eta H_0\eta^2)^{1/3}}\int_0^z\frac{dz'(1+z')^2}{[E(z')]^{7/3}}
\approx\frac{0.27 (f_C/0.1)(\Omega_m/0.3)}{(M_{500}h_{0.7}/10^{14}\msun)^{1/3}
(\eta/500)^{2/3}}\times \frac{2}{\Omega_m}\left\{1-\frac{1}{[E(z)]^{1/3}}\right\}
\nonumber\\
&\approx&\frac{0.06\Dgpc h_{0.7}^{2/3}(f_C/0.1)(\Omega_m/0.3)}{(M_{500}/10^{14}\msun)^{1/3}
(\eta/500)^{2/3}}~,
\label{intersections}
\ea
where the second line is for low $z$;
for a distribution of masses, we replace $M_{500}^{-1/3}$ by its mean, which may be skewed toward lower
masses, depending on the galaxy cluster mass distribution.
According to Eq. (\ref{intersections}), intersections
are rather likely (30\%) out to $z\simeq 1$, and approaches $\simeq 1$ at larger $z$,
 Moreover, intersections are likelier at smaller density contrasts:
$N_{\rm C, int}\propto M_\eta^{-1/3}\eta^{-2/3}$, and $M_\eta$ ultimately levels off as $\eta$ decreases.

The expected angular separation $\theta_1$ between a given line of sight and the nearest cluster
is 
\be
\theta_1=\left[\frac{8GM_CH_0}{f_C\Omega_mc^3\rtil^3(z)}\right]^{1/2}\approx
\frac{1.9\,{\rm arcmin}\,(M_Ch_{0.7}/10^{14}\msun)^{1/2}}{(f_C/0.1)^{1/2}(\Omega_c/0.3)^{1/2}[\rtil(z)]^{3/2}}
\approx\frac{17\,{\rm arcmin}\,(M_C/10^{14}\msun)^{1/2}}{(f_C/0.1)^{1/2}(\Omega_c/0.3)^{1/2}h_{0.7}\Dgpc^{3/2}}
\label{thetaone}
~,
\ee
where once again the last approximation is for low $z$, where $\rtil(z)\simeq z\simeq H_0d/c$.
Therefore, the probability that the line of sight to a particular FRB passes close to a cluster is fairly large
if FRBs may be seen out to $\Dgpc\simeq 1$.
This estimate ignores clustering of clusters, which enhances the
probability of seeing additional clusters near the direction to a particular cluster.

Various studies of the galaxy cluster mass function are consistent with $f_C\sim 0.05-0.1$, 
with a cut-off above $M_\star\sim 10^{14}\msun$.
\citep[e.g][]{{1996astro.ph.11148B},{1999fsu..conf..135B},{2002ApJ...567..716R},
{2007ApJ...657..183R},{2010MNRAS.407..533W},{2014A&A...566A...1T}}.  
The probability that an ERB originates either within or behind a cluster is therefore $\sim 10-15\%$ if
$\Dgpc\simeq 1$; the probability is $\sim 35-40\%$ for ERBs originating from $z\gtrsim 1$. 
These probabilities are high enough that some of the ERBs detected to date may have been located in or
behind clusters of galaxies; the DMs for such an ERB will have a considerable contribution from the cluster.
Thus, FRB redshifts and distances inferred under the assumption that DMs arise solely from a smooth
IGM are overestimates: actual redshifts and distances are likely to be smaller.

Because of the DM fluctuations introduced by clusters and their environs, it is not unlikely that two FRBs 
separated by $\sim 10$ arcmin on the sky have DMs that differ by $\sim 300\,{\rm cm^{-3}\,pc}$ 
\citep[see][and Eqs. (\ref{dmeta}) and (\ref{thetaone})]{2014ApJ...780L..33M}.
This may be the explanation for why FRBs 110220 and 140514, which are separated by about 9 arcmin on the sky,
have DMs that differ by about 380 ${\rm pc\,cm^{-3}}$; see \cite{2013Sci...341...53T} and \cite{2015arXiv150402165P}.
Although we do not offer any scenario in which it is likely that two FRBs are that close on the sky, given that
they are, the line of sight to FRB 110220 may pass through a cluster, whereas the line of sight to FRB 140514
passes outside. This unlikely coincidence of positions on the sky may therefore offer indirect evidence for the 
extragalactic origin of these FRBs, since the additional contribution from an intervening galaxy cluster 
accounts simply for their DM difference.

\section{Observational Tests}
\label{sec:tests}

We expect a number of  observational features from an  ERB source population comprising extragalactic NS, some of which constitute tests of the model.  
Some of these features in fact are generic to {\it any} extragalactic source population. 

\subsection{Temporal Substructure,  Polarization, and Faraday Rotation  of ERBs}

ERB pulses with ms durations are necessarily composed of sub-ns shot pulses, regardless of 
the nature of the emitting source.    Our analysis of the NS-magnetosphere case suggests that the 
number of shot pulses must be very large, indicating that ERBs analyzed with high time resolution
will be consistent with amplitude-modulated noise.  However, a nearby (low-z) population requires far fewer shot pulses, raising the prospect that individual  shot pulses or other substructure might be detectable. However, pulse broadening from extragalactic scattering  will prevent detection of shot pulses for at least some of the objects where scattering is seen.  Detection of individual shot 
pulses would imply that the underlying population of ERB sources is nearby on a cosmological scale and  would require that scattering broadening of pulses generally be small.  

While we expect a null result,  it  nonetheless is worthwhile 
to use radio telescope backend processors  with sub-$\mu s$ resolution and test whether 
the shot noise has a very high rate, as we conclude, or a low rate.     Polarization measurements 
may be informative about magnetic field conditions in ERB sources and whether depolarization occurs 
from incoherent summing of shot pulses and from propagation effects in the sources or along the 
line of sight.

ERB polarization also may indicate whether the ms-pulse duration is associated with rotation of a beam, as in radio pulsars, or due to true temporal modulation, which might yield  a constant polarization state. Single pulses from pulsars generally show elliptical polarization and a rotation of the plane of polarization across the pulse,   which is consistent with relativistic beaming from particles moving along curved magnetic field lines combined with the pulsar's spin.  However, individual shot pulses from the Crab pulsar show large degrees of circular polarization and changes in handedness  between neighboring shot pulses  that are argued to be due to temporal modulations \citep{2003Natur.422..141H}.   Linearly polarized ERBs can be used to measure Faraday rotation along the line of sight. If ERB sources are embedded in host galaxies, a wide range of rotation measures
$\vert \rm RM \vert \le 500 $~rad~m$^{-2}$ can be expected \citep[][]{2012A&A...542A..93O}.   Contributions to RM from intergalactic gas  have been identified \cite[][]{2006ApJ...637...19X, 2008ApJ...676...70K} that would be relevant to ERB sources outside spiral galaxies at high redshifts.

\subsection{Reobservations and Localization of ERB Lines of Sight}

Attempts to redetect ERBs from a particular object are problematic for any large source population,
such as extragalactic NS. 
In Section~\ref{sec:reobs1} we demonstrated that reobservations do not provide definitive constraints on the event rate per NS if the population extends to cosmological distances because many
NS  are within the instantaneous FoV of the  largest single-dish telescopes and, for distant populations,  even within the synthesized beam of array telescopes.     For example,  observations with the VLA in its A configuration  at 1.5~GHz provide a synthesized beam width $\sim 1$~arc sec that includes $\sim 2000d_{\rm Gpc}^3$~NS.  So long as ERBs from a given source are not bunched in time, it is equally likely for any
of these sources to emit a burst after an initial one is discovered.  One source per FoV requires either a VLB array with milli-arcsec resolution or a population with $\dmax \lesssim 80$~Mpc.  While the dispersion measure can be used to discriminate between sources,  if it is dominated by the intergalactic medium the many sources at similar redshifts will show similar values. 

Localization of ERB sources  would allow  discrimination between a
young NS population, which is expected to be associated with star-forming galaxies, and old objects which may have no association if they involve sources kicked out of galaxies.   However,  sources are not likely to be disassociated from galaxy clusters, so larger ERB samples can be tested for coincidence with cluster catalogs as well as galaxy catalogs.     Any galaxy-ERB associations combined with redetections in followup observations would allow one to argue statistically that the original ERB
source has been redetected even if there are many NS within the telescope FoV. 

\subsection{Interstellar Scintillation and Extragalactic Scattering}

Diffractive interstellar scintillations (DISS) are seen from Galactic pulsars as a consequence of multi-path propagation.   Fully-modulated (100\%) intensity variations in time and frequency require a source size  smaller than the isoplanatic angle,  $\theta_{\rm iso} \approx  \lambda/ D\theta_{\rm s, ISM}$,
where  $\theta_{\rm s, ISM}$ is the scattering diameter \citep[e.g.][]{2003ApJ...596.1142C}. 
Larger sources will attenuate DISS by a factor
$\theta_{\rm psr} /\theta_{\rm iso}$.  
For kpc distances and mas scattering angles, $\theta_{\rm iso} \approx 0.4~\mu {\rm as} \, \nu^{1.2} D_{\rm kpc}^{-1}$.  For lines of sight at high Galactic latitudes,  the distance $D$ is replaced by
the scale height $H_{\rm ISM} \approx 0.5$~kpc.    The NE2001 model 
\citep[][]{2002astro.ph..7156C} uses a distributed model for the scattering and yields
$\theta_{\rm iso} \approx  2~\mu as\, \nu^{1.2}$ at  a Galactic latitude $b = 90^{\circ}$.   
 
 Magnetospheres of Galactic pulsars easily satisfy 
$\theta_{\rm psr} \lesssim \theta_{\rm iso}$.  
Galactic pulsars have measured  characteristic time scales of seconds to hours
and characteristic bandwidths of tens of kHz to 100~MHz.  
ERB sources should show DISS if they are intrinsically small, as they must be from light-travel arguments, but only if they are not angularly broadened significantly by scattering along the 
extragalactic portions of their lines of sight.  We therefore consider extragalactic scattering in either the IGM,  a host galaxy, or  a galaxy cluster combined with scattering in the foreground ISM.      
Temporal scintillations of ERBs cannot be measured in any case because the burst duration is short.  However, frequency structure in the spectrum is potentially measurable.   

Several ERBs show pulse asymmetries with the strong frequency dependence indicative of plasma scattering.  Given their high Galactic latitudes, this scattering must be due to extragalactic scattering.   We use the scattering time $\tau_s$ to estimate the scattering diameter and compare it to the isoplanatic angle for DISS. 
For a thin scattering screen at distance $D_s$ from the source, the pulse broadening time is 
$\tau_s =  g_s D\theta_o^2 / 2c$, where  $\theta_o$ is the scattered source size and 
$g_s = D/D_s-1$ takes into account the scattering geometry. 
ERB scattering times  $\sim 1~$ms  imply 
$\theta_o\approx\sqrt{2c\tau_s/g_s D} \approx  30~{\rm\mu as}\, \sqrt{\tau_{\rm ms}/g_s D_{\rm Gpc}}$.

First consider scattering midway along the LOS, $D_s = D/2$, which gives $g_s = 1$.   The scattering diameter `seen' by the foreground ISM is then too large to show DISS, particularly  for sources
nearer than 1~Gpc.    No frequency structure is expected in the spectrum for this reason and also 
because its characteristic bandwidth would be $\Delta\nu_{\rm DISS} \approx 1/2\pi\tau_s \approx 160$~Hz. 

However,  extragalactic scattering close to the source with $D_s \ll D$ yields  $g_s \gg 1$ and
 a much smaller scattered source diameter. For $D_s \sim 1$~Mpc, as might be appropriate for a source scattered by ionized gas in an ERB's host cluster, 
$
\theta_o \approx  0.9~{\rm \mu as}\, \sqrt{\tau_{\rm ms} D_{\rm s, Mpc}}/D_{\rm Gpc}.
$
which is small enough to allow fully-modulated DISS.  Sources embedded in and scattered by a host galaxy with $D_s\sim 1$~kpc imply even smaller source sizes (by a factor of 30). 

When diffractive interstellar scintillations are expected,  the frequency structure will have a characteristic scale given by $\Delta\nu_{\rm DISS} \sim 1 / 2\pi \tau_{\rm s, ISM}$ where
$\tau_{\rm s, ISM}$ is the pulse broadening time expected from  scattering in the Milky Way. 
For $b = 90^{\circ}$, a scintillation bandwidth $\Delta\nu_{\rm DISS} \sim 4$~MHz is expected.
Scintillation structure on frequency scales of this order can be searched for and would demonstrate
that extragalactic scattering occurs close to the source.  Absence of frequency structure of course would imply that scattering is more distributed along the line of sight.

\subsection{Giant Pulse Energetics}

Much of our discussion has involved the energetics of radio bursts that refer to previously observed
giant pulses from the Crab pulsar.   It would be highly informative to monitor the Crab pulsar over
thousands of hours, much longer than the longest dwell times to date, in order to identify or constrain a maximum in the distribution of pulse amplitudes.     Any such maximum for the Crab pulsar may signify similar maxima in the distributions of ERB amplitudes from particular sources and from the ensemble of sources.

\section{Summary \& Conclusions}
\label{sec:conclusions}

In this paper we have examined the hypothesis that fast radio bursts are associated with
rare,  bright pulses from extragalactic neutron stars.   Our motivation is twofold. First, NS are known to emit bright pulses with a wide range of durations, including the ms widths of ERBs,
so associating fast radio bursts with extragalactic NSs involves extrapolation from known phenomena.     Second, the large number of NS that exist in a Hubble volume can easily produce the inferred ERB rate even if only a single burst is produced in the lifetime of each NS. An added virtue, however,  is that multiple bursts per object allow the population to be closer and relax the energy requirements on the radiation process. The multiplicity can still be small enough that no individual source would repeat over time scales of months to even centuries. However, the source population could be near enough for repeats to be observed on human time scales.   By contrast, burst interpretations involving one-off events per object, such as evaporating black holes or from stellar implosions, necessarily require a large cosmological volume to account for the ERB rate.  Finally, even if NS do not turn out to be the source population, we consider it a matter of due diligence to consider NS prior to invoking more exotic explanations. 

We have found that the radiation mechanism is highly constrained.   Independent of the source population,  
sub-ns structure in the radiation field must be produced through a coherent process and the overall 
ms widths of ERBs require incoherent superposition of a large number of individually coherent shot pulses.  
We have specifically examined curvature radiation in conditions similar to the magnetospheres of active pulsars.  
Such radiation can account for the lone, 2~MJy shot pulse from the Crab pulsar measured by 
\citet[][]{2007ApJ...670..693H} via coherent radiation from a nearly neutral clump, with net charge 
$\sim10^{21}e$ and total lepton number $\gtrsim 10^{27}$, near the light cylinder.
The radiation strength of the Crab shot pulse, measured as fluence times distance squared, 
$\sim 1$~MJy-ns-kpc$^2 = 1$~Jy-ms-kpc$^2$.  ERBs have  strengths
$\sim 10^{12}d_{\rm Gpc}^{\,2} $~Jy-ms-Gpc$^2 $.  Such large radiation strengths cannot be due to single coherent regions
because the ms time scales require a temporal spread of shot-pulse arrival times.  
If an ERB comprises numerous  shot pulses that individually are of order the largest seen from the Crab  pulsar, the number  involved is $\sim 10^{12}\Dgpc^2$,
which is sensitive to distance but very large at all extragalactic distances.
However, we note that large numbers of shot pulses comprise single pulses from most pulsars, which have durations up to 100s of ms.  Comparing the total energy
available to fuel an outburst from a NS magnetosphere to the total energy required observationally yields a maximum
source distance; for an outburst originating near the light cylinder of a Crab-like NS ($P\approx 30$ ms, $\dot P/P
\approx (1000\,{\rm years})^{-1}$) we find $\Dgpc\lesssim 0.1$. 
Bursts from ordinary pulsars are similarly constrained to arise from local group galaxies but  in some circumstances (namely large magnetic fields and low emission altitudes) could be seen from further away.  Bursts from magnetars could similarly originate from the local group and perhaps further. 
These rough distance bounds can be extended 
to $\Dgpc\simeq 1$ if outbursts underlying ERBs are associated with young NS (large $\dot P/P$ and small $P$), 
especially if the outburst occurs well  within the light cylinder,  
where the magnetic field and available magnetospheric energy reservoir are larger.    However, there may be a lower bound 
on the emission altitude $\sim c W_i$  based on measured ERB widths $W_i\sim$ ms, and the time needed for coherence to be established.

A triggering mechanism is needed for infrequent, high amplitude bursts from NS (or from any source that repeats).  For emission from a magnetosphere, the trigger could either be internal, such as from an extraordinarily large rotation-altering
event (e.g. a spin glitch), or external, such as infalling debris
\citep[e.g.][]{2008ApJ...682.1152C} that stimulates a pair cascade or from a large reconnection event in NS magnetotails  
that injects particles and energy into the magnetosphere.  Internal triggers may inject more energy than can be carried 
in by external material or charges.   However, external triggers may indicate an important role for the local 
environment of ERB sources.  Environmental effects, in turn, have implications for the relative contributions 
to ERB dispersion measures.

DMs of ERBs have largely been attributed to the IGM combined with a foreground contribution from the Milky Way 
and a similar amount from the host galaxy\footnote{We note that a few authors have argued that ERBs are 
in fact Galactic sources.}, yielding relatively high redshifts derived from the required, substantial IGM contribution.  
Generally, an intervening galaxy or host or intervening galaxy cluster will also make contributions, as will 
any dense gas local to the source itself.   We note that the 
relatively large redshifts attributed to ERBs ($0.5\lesssim z \lesssim 1$) yield a non-negligible probability 
for the line of sight encountering a galaxy cluster; additionally, some ERB sources may reside  within clusters.  
Since clusters can account for a large fraction of the measured DMs, ignoring them is inconsistent 
with the attributed redshifts for at least some of the ERBs.  If the local environment plays a role 
in triggering ERBs, the underlying sources may reside in dense regions within host galaxies, therefore 
providing another sizable contribution to DMs.   An extreme case would be triggering by jet particles 
from supermassive black holes in galactic centres, which could account for most of the measured DMs 
and allow the population to reside at small $\dmax$. 

As others have stated \citep[e.g.][]{2015ApJ...807...16L}, the most valuable new observations will be 
the  localization of ERB sources through high time resolution radio imaging.  These will allow a 
determination of the distance scale and provide constraints on the contributions of host galaxies 
and the IGM to dispersion measures. 
Our analysis suggests that it is unlikely that X-and-$\gamma$-ray emission will be seen from ERBs 
if the ratio of high energy emission to radio emission scales like that from pulsars.

We thank Shami Chatterjee and  Tim Hankins for useful conversations and Jean Eilek and Tim Hankins for providing the data used in Figure 1.  We also thank the anonymous referee for several suggestions that improved the paper significantly. 
This work was supported by NSF Grant Phys 1104617 and NASA grant  NNX13AH42G.

\newpage
\appendix

\section{Shot Noise with Variable Rate and Amplitude}
\label{app:flux}
\label{app:ramsn}

\newcommand{\nuz}{\nu_0}

Consider a (scalar) shot pulse $E(t)$  with duration $W_s \lesssim 1$~ns whose  Fourier transform 
$\tilde E(\nu)  = \int dt\, E(t) e^{-2\pi i\nu t}$  is sampled with a
receiver with bandwidth $\Br$ centred on a frequency $\nuz$.    The selected
electric field for a rectangular bandpass shape is 
\be
E(t, \nuz;\Br) &=& \int_{\nuz-\Br/2}^{\nuz+\Br/2} \!\!\!\!\!\! d\nu\, \tilde E(\nu) e^{+2\pi i \nu t} 
+ \,\, \text{CC},
\ee
where the negative frequencies correspond to  the complex conjugate of the first term. 
 For an unresolved pulse  with an intrinsic bandwidth $\sim W_s^{-1} \gg \Br$, we have
\be
E(t, \nuz;\Br) &\approx& \Br \tilde E(\nuz) e^{+2\pi i \nuz t} \,\sinc\ \Br t + \,\, \text{CC},
\label{eq:shot}
\ee
where $\sinc~ x = (\sin \pi x) / \pi x$.  This equation will be approximately valid even for 
$W_s^{-1} \gtrsim \Br$.  In practice, the signal is usually heterodyned to zero frequency (baseband)
through multiplication by  $e^{-2\pi i \nuz t}$ to obtain the complex pulse 
$ \Br \tilde E(\nuz)  \,\sinc\ \Br t$.   Dispersion in ionized gas is not included in the model because it is assumed to be removed as one of the operations made on the baseband signal.  Scattering is also excluded here though it will cause individual shots to overlap and is a factor in the assessment of  total ERB widths.

The power collected in a unit area is $\sim (c /2\pi) \vert E(t, \nu;\Br) \vert^2$ so the measured
flux density of a single shot is
\be
\sshotz(t) = \frac{c \Br\vert \tilde E(\nuz) \vert^2\,\sinc^2\Br t} {2\pi}
	\equiv \sshotzmax\,  \sinc^2\Br t~.
\label{eq:Snu2}
\ee 
The measured shot pulse and  its maximum $\sshotzmax$ 
both depend on the receiver bandwidth but the fluence $\ashotz$ (the integrated flux density) is independent of the bandwidth because $ \Br \int dt\, \sinc^2 \Br t = 1$, 
\be
\ashotz = \int dt\, \sshotz(t) =
\frac{\sshotzmax} {\Br}
=
 \frac{c \vert \tilde E(\nuz) \vert^2}{2\pi}.
\ee

An incoherent superposition of coherent shot pulses  accounts for the 
electromagnetic and statistical properties of observed pulses  with  ms durations, as described in the main text.   Superposing shot pulses like  Eq.~\ref{eq:shot}, each having width $\sim \Br^{-1}$, arrival time $t_{s_j} $, and amplitude $\Br \tilde E_j(\nu)$ ,  the electric field is 
\be
E(t,\nu,\Delta\nu_r)= \Br \sum_j{\tilde E}_j(\nu)e^{2\pi\nu(t-t_{s_j})}\,\sinc\,\Delta\nu_r(t-t_{s_j}),
\ee
where we have dropped the subscript on the center frequency. 

When the number of shots is large,   the total flux density is the sum of terms like Eq.~\ref{eq:Snu2},    
\be
S_\nu(t) \approx \sum_j  {\sshotmax}_j \, \sinc^2 \Br(t-t_{s_j}) ;
\label{eq:Snusum}
\ee
cross terms between shot pulses  are negligible because  phases associated with the arrival times $t_{s_j}$ are  assumed statistically independent.
Using the characteristic function of a Poisson process \citep{Rice44, Papoulis91}, the ensemble average  depends on  the mean of the product of the shot pulse rate $\eta_s(t)$ and the  shot amplitude $\sshotmax(t)$,
\be
\langle S_\nu(t) \rangle  \approx
	\int d\tp\, \left\langle \eta_s(\tp)  \sshotmax(\tp)\right\rangle \,  \sinc^2\Br (t - \tp)
	\approx \left\langle \eta_s(t)   A_{\nu}(t)\right\rangle,
\label{eq:Snumean}
\ee
where we have defined  the  fluence per shot, $ A_{\nu}(t)   =  \sshotmax(t) / \Br$.
The second equality follows because
the composite  pulse is typically   much broader than $\Br^{-1}$, requiring that the rate and mean amplitude vary slowly compared to the $\sinc^2$ function. 

Individual pulses  have the same form as the ensemble average in Eq.~\ref{eq:Snumean} but are noisy from the finite number of shot pulses.   Factoring the rate and fluence for simplicity, the flux density is
\be
S_{\nu}(t) = \eta_s(t) A_\nu(t). 
\ee 
An ERB comprising  $N_i$ incoherently summed shot pulses over a width $W_i$  has a mean rate 
$\eta_s \simeq N_i / W_i$.   Alternative forms for the {\it peak}  flux density of the ERB  are then
\be
S_{\nu, \rm max} 
		\approx  
		\sshotmax  \left(\frac{\eta_s }{\Br}\right) 
		\approx N_i \sshotmax \left(\frac{W_r }{W_i}\right),
\label{eq:Snuapprox}
\ee 
where the time scale $W_r = 1/\Br$.  The first expression says that the peak ERB flux density is larger than the  peak single-shot flux density  by the ratio of the shot rate to the bandwidth, which can be large. The second says that the ERB peak is equal to the total flux density in all the shots diluted by the
factor $W_r / W_i \ll 1$.   

The expressions given so far use the `native' resolution of the receiver bandwidth, which will be less than 1~$\mu s$, typically.  Data are usually smoothed over many such resolution elements to a time scale $\tau_{\rm smooth} $ that is somewhat smaller than the actual pulse width $W_i$
to approximate matched-filtering detection.   In this case we would make the replacements $W_r = 1/\Br \to \tau_{\rm smooth}$ and $\sshotmax \to \sshotmax \times W_r / \tau_{\rm smooth}$, leaving the peak ERB flux unchanged in the mean, as expected.

\def\ttil{{\tilde t}}
\def\omtil{2\pi\nutil}
\def\nutil{{\tilde\nu}}
\def\Evtil{{\mbox{\boldmath ${\tilde E}$}}}
\def\Svec{{\mbox{\boldmath $S$}}}
\def\Bvec{{\mbox{\boldmath $B$}}}

\def\tec{t_{e,c}}
\def\temc{\te-\tec}
\def\omtemc{\Omega(\temc)}
\def\dtheta{\Delta\theta}
\def\dphi{\Delta\phi}
\def\dr{\Delta r}
\def\drl{\Delta R_l}
\def\eps{\epsilon}
\def\lhat{{\mbox{\boldmath $\hat\ell$}}}
\def\nuunit{\nu_{\rm GHz}}
\def\rlc{r_{lc}}
\def\omco{\omega_{coh}}
\def\nucounit{\nu_{coh,{\rm GHz}}}
\def\Dtec{\Delta\tec}
\def\epsco{\eps_{coh}}
\def\kapco{\kappa_{coh}}

\section{Conditions for Coherence}
\label{app:Coherence}

Let the observer be in direction $\nhat$ relative to the the source, and let $\tec$ be the time when emission
is optimally beamed toward the observer, for a given accelerated charge moving on a path
with radius of curvature $r_c$. This is the time when $\kappa=1-\nhat
\dotprod\vvec(t_e)$ is smallest; for strong beaming, this happens when $\nhat\dotprod\avec(t_e)=0$ (i.e. when
the acceleration is perpendicular to the line of sight), which is not guaranteed to occur but is required for super-bright
emission. At this time, the charge is at $\rvec(\tec)$. 
The electromagnetic field at an observation point $(t,\rvec)$ depends on $c(t-\tec)-\nhat\dotprod[\rvec-\rvec(\tec)]$.
For example, for curvature radiation from a charge $q$ moving along an instantaneously circular path with curvature
$\chat r_c$, the electric field for an observer at $\nhat d$ is
\be
\frac{dr_c\Evec(t,\nhat d)}{q}\approx\frac{\chat(1-\psi_e^2)+2\chat\crossprod\nhat\psi_e\eps/\sqrt{2\kappa_0}}
{\kappa_0^2(1+\psi_e^2)^3}
\ee
where 
$\psi_e=(\sqrt{\psi^2+1}+\psi)^{1/3}-(\sqrt{\psi^2+1}-\psi)^{1/3}$
with $\psi=(3/2r_c\kappa_0^{3/2})[ct-d-(c\tec-\nhat\dotprod\rvec(\tec))]$ and $\kappa_0=1-v\cos\eps\simeq
\onehalf(\eps^2+1/\gamma^2)$. Geometrically, the observer is at angle $\eps$ above the plane of the 
(instantaneously circular) orbit of the emitting particle.
The electric fields for a collection of particles with a range of peak emission times and positions
$\Delta\tec$ and $\Delta\rvec(\tec)$ add coherently at frequencies below $\omco$ if
\be
\vert\Delta[c\tec-\nhat\dotprod\rvec(\tec)]\vert\lesssim r_c(2\kapco)^{3/2}= r_c\left(\epsco^2+\frac{1}{\gamma^2}\right)^{3/2}
\equiv\frac{c}{\omco}~,
\label{coherencesummary}
\ee
where 
$\vert\eps\vert=\vert\epsco\vert$ at $\omco$.

Eq. (\ref{coherencesummary}) is derived by considering how long a pulse from a given charge is bright {\sl as seen
by the observer}.
Eq. (\ref{coherencesummary}) has a simple interpretation: at the observer, sharply pulsed electric fields will add
up coherently if they overlap. At a given frequency $\omega$ in a Fourier decomposition, the observer detects 
a superposition of electric fields with different phases;
the phase associated with reaching optimal beaming at time $\tec$ is $\omega[c\tec-\nhat\dotprod\rvec(\tec)]$, 
so $\omega[c\Delta\tec-\nhat\dotprod\Delta\rvec(\tec)]$ are the relative phases
associated with a set of waves that reach optimal beaming at different times and places, and these waves add
coherently for relative phases with magnitudes $\lesssim 1$.
Note that the volumes constrained by Eq. (\ref{coherencesummary}) represent {\sl upper bounds}: actual volumes
that emit coherently will depend on the emission mechanism.
For example,
for charges that are all launched at the same place but at different times and travel along identical trajectories,
hence reach optimal beaming at the same place,
Eq. (\ref{coherencesummary}) implies that the range of optimal beaming times is $\Delta\tec\lesssim r_c(2\kapco)^{3/2}$.

As an example,
 we consider motion along magnetic dipole field lines,  
$r=R_l\sin^2\theta$.
For a dipole field with symmetry axis along $\zhat$ the unit vector tangent to a field line is
\be
\that=\frac{3\cos\theta\rhat-\zhat}{\sqrt{3\cos^2\theta+1}}=\frac{3\cos\theta\sin\theta(\xhat\cos\phi+\yhat\sin\phi)
+(3\cos^2\theta-1)\zhat}{\sqrt{3\cos^2\theta+1}}
\ee
for angles $\theta, \phi$ referenced to the dipole axis
and the unit vector in the curvature direction is 
\be
\chat=\phihat\crossprod\that=\frac{(3\cos^2\theta-1)(\xhat\cos\phi+\yhat\sin\phi)-3\cos\theta\sin\theta\zhat}
{\sqrt{3\cos^2\theta+1}}~.
\ee
Using $d\that/ds=\chat / r_c$, the radius of curvature of a field line is 
\be
r_c
=\frac {r(3\cos^2\theta+1)^{3/2}} {3\sin\theta(\cos^2\theta+1)}~.  
\ee
The direction to the observer is 
$\nhat=\xhat\sin\alpha + \zhat\cos\alpha$. At the time of optimal beaming from a particular field line
$\nhat\dotprod\avec=0=\nhat\dotprod\chat$, since the acceleration $\avec$ is along $\chat$. At optimal
beaming for a particular field line, $\that\dotprod\nhat=\cos\epsilon\simeq 1-\onehalf\epsilon^2+\cdots$;
we define a reference field line to be the one for which $\epsilon=0$, and $\theta=\theta(0)$ and $\phi=0$. For any
other point, optimal beaming occurs when $0=-3\cos\theta\sin\theta\cos\alpha+(3\cos^2\theta-1)\sin\alpha
\cos\phi$; we then find a displacement $\Delta\rvec$ whose components are
\ba
\nhat\dotprod\Delta\rvec&=&\sin\theta(0)\left[\frac{2\drl\sin\alpha}{3}-\frac{\epsilon^2R_l\cos\alpha(\sqrt{8+\cos^2\alpha}
+\cos\alpha)^2}{8\sin\alpha\sqrt{8+\cos^2\alpha}}\right]
\nonumber\\
\chat(0)\dotprod\Delta\rvec&=&-\frac{\drl(\sqrt{8+\cos^2\alpha}-3\cos\alpha)}{6}
\nonumber\\
\phihat\dotprod\Delta\rvec&=&\pm\frac{\epsilon R_l\sin^3\theta(0)}{\sin\alpha}~,
\label{displacements}
\ea
where the contributions $\propto R_l$ come from either remaining on a field line but sliding along it or rotating
azimuthally to another field line with the same $R_l$, and the contribution $\propto\drl$ comes from moving to a 
different (set of) field line(s). The coherence condition Eq. (\ref{coherencesummary}) 
only involves the first component of Eq. (\ref{displacements}), but this condition constrains the other components 
indirectly.

First consider what happens when $\Dtec=0$ i.e. find the volume when the coherent emission is simultaneous at the
source. In that case, we can use Eq. (\ref{coherencesummary}) to find $\drl/R_l$; unless $\eps\ll\gamma^{-3/2}\ll\gamma^{-1}$
we get $\drl/R_l=\eps^2 f_{\nhat}(\cos\alpha)$, and therefore $\chat(0)\dotprod\Delta\rvec=\eps^2r_c f_{\chat}(\cos\alpha)$,
where the functions are found by combining Eq. (\ref{displacements}) and (\ref{coherencesummary}); 
$\phihat\dotprod\Delta\rvec=
\pm\epsilon r_c f_{\phihat}(\cos\alpha)$ is already determined by Eq. (\ref{displacements}). We therefore get an 
emitting area $\mathscr{A}_{\rm c}=\vert\eps^3\vert\ r_c^2f_{\mathscr{A}_{\rm c}}(\cos\alpha)=(cr_c/\omco)
f_{\mathscr{A}_{\rm c}}(\cos\alpha)$, and an emitting volume $\Vc=2c/\omco\times\mathscr{A}_{\rm c}$ for an individual coherent pulse; explicitly
\be
\Vc
\lesssim\frac{64c^2r_c\sin\alpha\cos\alpha\sqrt{8+\cos^2\alpha}}
{3\omco^2(\sqrt{8+\cos^2\alpha}+\cos\alpha)(\sqrt{8+\cos^2\alpha}+3\cos\alpha)^2}
\label{volume}
\ee
where Eq. (\ref{coherencesummary}) implies $\vert\eps\vert\lesssim(c/\omco r_c)^{1/3}$ 
and $\vert\nhat\dotprod\Delta\rvec\vert\lesssim c/\omco$.
Numerically, the characteristic upper limit on the 
volume for a set of charges emitting coherently with simultaneous $\tec$ are
\be
\frac{c^2r_c}{\omco^2}=\frac{1.09\times 10^{11}\,{\rm cm^3}\,(r_c/\rlc)P}{\nucounit^2},
\label{charareavol}
\ee
where the curvature radius has been expressed in units of  $\rlc  = cP/2\pi$, 
the light cylinder radius for spin period $P$. Note that $\Vc\sim(\omco r_c/c)(c/\omco)^3\sim\eps^{-3}(c/\omco)^3
\sim\gamma^3(c/\omco)^3\gg(c/\omco)^3$.
The thickness of the volume along $\nhat$ is $2c/\omco$, so the coherent volume is severely flattened,
with a cross-sectional area $\mathscr{A}_{\rm c}\sim  cr_c/\omco$. The area is irregular, with a length $\sim r_c/\gamma^2
\sim\gamma c/\omco$
along $\chat$ and $\sim r_c/\gamma\sim\gamma^2 c/\omco$ along $\nhat\crossprod\chat$.

Note that this irregular shape arises from considering the largest possible
region that can emit coherently, allowing the shape to be determined by
Eq. (\ref{coherencesummary}) and the magnetic field geometry. If the shape
of the region emitting coherently is {\sl prescribed} then it must fit
entirely inside the largest possible coherently emitting region. For example,
if we insist that coherent emission arises from a spherical region, then
the radius of the sphere must be $\leq c/\omco$, and the volume of the
coherently emitting sphere is $\lesssim (c/\omco)^3$, as is often assumed.
However, if the shape is not constrained to be regular, Eq. (\ref{charareavol})
permits significantly larger coherently-emitting volumes.

The $\Dtec=0$ limit is valid for $\Dtec\ll \eps^2r_c/c$.
For $\Dtec\gtrsim \eps^2r_c/c\sim (\omco\eps)^{-1}\sim\gamma/\omco$, the time differences for optimal beaming are
sufficiently non-simultaneous to affect the coherence volume substantially: the coherence condition is satisfied when 
$c\Dtec\simeq\nhat\dotprod\Delta\rvec$, or
\be
\drl\simeq\frac{3c\Dtec}{2\sin\theta(0)\sin\alpha}\,.
\label{drldtec}
\ee
Eq. (\ref{drldtec}) means that
if the burst  lasts a time $\Dtec\gtrsim\eps^2r_c/c$ the point where emission is beamed right at the
observer shifts through a succession of values of $\drl$. 
We then find 
\ba
\Delta\rvec\dotprod\chat(0)&\simeq&\frac{2c\Dtec\sin\alpha}{\sqrt{8+\cos^2\alpha}+3\cos\alpha}
\nonumber\\
\Delta\rvec\dotprod\phihat&\simeq&\pm\frac{\eps R_l\sin^3\theta(0)}{\sin\alpha}~;
\ea
the volume of the coherently emitting particles is
\be
\Vc\lesssim\frac{4(c/\omco r_c)^{1/3} r_c(c\Dtec)^2\sqrt{8+\cos^2\alpha}(\sqrt{8+\cos^2\alpha}-3\cos\alpha)}
{3(\sqrt{8+\cos^2\alpha}+\cos\alpha)(\sqrt{8+\cos^2\alpha}+3\cos\alpha)}
\label{vcohdtec}
~.
\ee
Eq. (\ref{vcohdtec}) differs from Eq. (\ref{volume}) by a factor $\sim (\omco\Dtec)^2\gamma^{-1}\gtrsim\gamma$.
The shape of the coherent region is irregular, but $\vert\nhat\dotprod\Delta\rvec\vert\sim\vert\chat(0)\dotprod
\Delta\rvec\vert\sim c\Dtec$ in this case, while $\vert\phihat\dotprod\Delta\rvec\vert\sim r_c/\gamma$.
If $\Dtec\sim r_c/c\gamma=P(r_c/\rlc)/2\pi\gamma=0.16P(r_c/\rlc)\gamma_3^{-1}$ ms 
then the extent of the region is similar in all three directions.

We have focussed on dipole field geometry in this section, but we expect many of the scalings to remain valid if
$r=R_l\mathscr{F}(\sin\theta)$, where $\mathscr{F} = \sin^2\theta$ for a dipolar field; that is, although the dependences on $\alpha$ will differ, if the field line
shapes only depend on a single length scale, then similar results should be obtained. Realistically, though, the shapes of pulsar magnetic field lines will depend on at least two lengths,
$R_l$ and the light cylinder radius $r_c$, and the situation is more complicated. However, to the extent that
only localized regions are involved in coherent emission, and assuming that field lines may be approximated as
circular locally, the basic results we found for the extent of coherent regions ought to be similar to what 
we found here for dipole geometry. 

\newpage

{

}

\end{document}